\documentclass[twocolumn,usenatbib]{aastex61}
\usepackage{natbib}

\newcommand{\KN}{AT2017gfo}
\newcommand{\cosmo}{$H_0 = 67.8$~km~s$^{-1}$~Mpc$^{-1}$, $\Omega_M = 0.308$ and $\Omega_{\Lambda} = 0.692$ \citep{PlanckCollaboration16}}

\received{}
\revised{}
\accepted{\today}

\submitjournal{ApJ}

\shorttitle{Diversity of kilonova emission in short gamma-ray bursts}
\shortauthors{Gompertz et al.}

\begin{document}

\title{The diversity of kilonova emission in short gamma-ray bursts}
\correspondingauthor{Benjamin Gompertz}
\email{b.gompertz@warwick.ac.uk}

\author{B. P. Gompertz}
\affil{Department of Physics, University of Warwick, Coventry, CV4 7AL, UK}

\author{A. J. Levan}
\affiliation{Department of Physics, University of Warwick, Coventry, CV4 7AL, UK}

\author{N. R. Tanvir}
\affiliation{Department of Physics and Astronomy, University of Leicester, Leicester, LE1 7RH, UK}

\author{J. Hjorth}
\affiliation{Dark Cosmology Centre, Niels Bohr Institute, University of Copenhagen, Juliane Maries Vej 30, 2100 Copenhagen, Denmark}

\author{S. Covino}
\affiliation{INAF, Osservatorio Astronomico di Brera, Via E. Bianchi 46, I-23807 Merate (LC), Italy}

\author{P. A. Evans}
\affiliation{Department of Physics and Astronomy, University of Leicester, Leicester, LE1 7RH, UK}

\author{A. S. Fruchter}
\affiliation{Space Telescope Science Institute, 3700 San Martin Drive, Baltimore, MD 21218, USA}

\author{C. Gonz\'alez-Fern\'andez}
\affiliation{Institute of Astronomy, University of Cambridge, Madingley Road, Cambridge, CB3 0HA, UK}

\author{Z. P. Jin}
\affiliation{Key Laboratory of dark Matter and Space Astronomy, Purple Mountain Observatory, Chinese Academy of Science, Nanjing, 210008, China}
\affiliation{School of Astronomy and Space Science, University of Science and Technology of China, Hefei, Anhui 230026, China}

\author{J. D. Lyman}
\affiliation{Department of Physics, University of Warwick, Coventry, CV4 7AL, UK}

\author{S. R. Oates}
\affiliation{Department of Physics, University of Warwick, Coventry, CV4 7AL, UK}

\author{P. T. O'Brien}
\affiliation{Department of Physics and Astronomy, University of Leicester, Leicester, LE1 7RH, UK}

\author{K. Wiersema}
\affiliation{Department of Physics, University of Warwick, Coventry, CV4 7AL, UK}

\begin{abstract}
The historic first joint detection of both gravitational wave and electromagnetic emission from a binary neutron star merger cemented the association between short gamma-ray bursts (SGRBs) and compact object mergers, as well as providing a well sampled multi-wavelength light curve of a radioactive kilonova (KN) for the first time. Here we compare the optical and near-infrared light curves of this KN, \KN{}, to the counterparts of a sample of nearby ($z < 0.5$) SGRBs to characterize their diversity in terms of their brightness distribution. Although at similar epochs \KN{} appears fainter than every SGRB-associated KN claimed so far, we find three bursts (GRBs 050509B, 061201 and 080905A) where, if the reported redshifts are correct, deep upper limits rule out the presence of a KN similar to \KN{} by several magnitudes. Combined with the properties of previously claimed KNe in SGRBs this suggests considerable diversity in the properties of KN drawn from compact object mergers, despite the similar physical conditions that are expected in many NS-NS mergers. We find that observer angle alone is not able to explain this diversity, which is likely a product of the merger type (NS-NS versus NS-BH) and the detailed properties of the binary (mass ratio, spins etc). Ultimately disentangling these properties should be possible through observations of SGRBs and gravitational wave sources, providing direct measurements of heavy element enrichment throughout the Universe.
\end{abstract}


\section{Introduction} \label{sec:intro}
Short gamma-ray bursts (SGRBs) have long been thought to be the products of the mergers of compact objects \citep{Rosswog03,Belczynski06,Nakar07} - either binary neutron star (BNS) or neutron star-black hole (NSBH) systems. In this framework, the energy release of the merger launches relativistic jets that produce $\gamma$-rays in internal shocks. A broad-band synchrotron afterglow, with emission ranging from X-ray to radio frequencies, is then produced as the outflow decelerates in the circumstellar environment \citep{Meszaros93b}. Such a merger is also expected to produce a faint optical/nIR transient known as a `kilonova' (KN; or `macronova') \citep{Li98,Rosswog05,Metzger10b} as ejected material rich in neutrons forms heavy elements through rapid neutron capture (r-process) nucleosynthesis \citep{Lattimer74,Eichler89,Freiburghaus99} that subsequently decay radioactively. However, the discovery of GW~170817 \citep{Ligo17a,Ligo17} by the Advanced Laser Interferometer Gravitational-Wave Observatory (LIGO) and Advanced Virgo provided the first direct evidence that the merger of a BNS has occurred. With it came the detection of SGRB\,170817A by \emph{Fermi} \citep{Fermi17,Goldstein17,vonKienlin17} and later INTEGRAL \citep{Savchenko17}. In broad-band follow-up observations \citep{Ligo17}, an optical transient \citep{Coulter17,Coulter17a,Soares-Santos17} identified as a KN,  \citep[\KN{}; e.g.][]{Cowperthwaite17,Chornock17,Evans17,Nicholl17,Pian17,Smartt17,Tanvir17} was detected, thus confirming the compact object merger origin for SGRBs and KNe.

Since 2005, the rapid localisations provided by the \emph{Swift} satellite have meant that a catalogue of SGRB afterglows at multiple wavelengths has been established. These afterglows are at their brightest just after the gamma-ray emission, and decay away in the minutes to days following the burst. Interestingly, in GW~170817 no afterglow was detected after the \emph{Fermi} trigger. Despite being tied to NGC 4993, a nearby early-type galaxy at $z = 0.009783$, or just $42.5$~Mpc \citep{Blanchard17,Hjorth17,Levan17}, no X-ray emission was detected down to $2.7\times10^{-13}$~erg~s$^{-1}$~cm$^{-2}$ at $0.62$ days post-trigger \citep{Evans17}. However, GRB jets are brightest on axis, and are expected to be fainter, and exhibit afterglows which rise later, as the observer moves off-axis. It is therefore tempting to ascribe the faint GRB \cite{Fermi17,Goldstein17,vonKienlin17}, and slowly rising afterglow emission \citep{Troja17,Mooley18,Lyman18,Margutti18} to an SGRB viewed away from the axis of its relativistic jet. Indeed, the GW data suggests a viewing angle of up to $28$ degrees from the rotation axis of the binary \citep{Evans17,Haggard17,Ligo17,Margutti17,Tanvir17,Mandel18}, consistent with an off-axis model. However, the widely adopted simplification of a `top hat' jet, in which emission drops off sharply outside of narrow beam, likely needs to be modified to include structure away from the core of the jet \citep[e.g.][]{Lazzati17}. Alternative models featuring emission from a mildly relativistic cocoon of ejecta have also been proposed \citep[e.g.][]{Kasliwal17b,Gottlieb18,Mooley18}, though these too may feature a polar jet that could be detected as a more typical cosmological SGRB if it had been oriented towards the Earth.

Due to their typical discovery via detection of the $\gamma$-ray jet, SGRBs are normally viewed down the jet axis where the afterglow is brightest and therefore most likely to mask a KN, which is expected to be a more isotropic component, that would normally be easier to see against the glare of the SGRB afterglow at angles away from the SGRB jet \citep{Metzger12}. None-the-less, candidate KNe have been observed in just a handful of events: the first and best claim coming with SGRB 130603B \citep{Tanvir13,Berger13}, followed by a re-analysis of SGRBs 060614 \citep{Yang15} and 050709 \citep{Jin16}, and a so far inconclusive fourth candidate in SGRB 160821B \citep{Jin17,Kasliwal17}. 

Through association with BNS mergers and KNe, SGRBs are now thought to be the site of a sizeable fraction of heavy element production in the Universe \citep{Lattimer74,Rosswog98,Goriely11,Korobkin12,Just15}, supplementing the apparently insufficient yields predicted for supernovae \citep{Thielemann11}. The identification of a clear KN signature accompanying GW~170817 and GRB~170817A provides the perfect opportunity to assess their general detectability in SGRBs and the variability in their properties. Known BNS in the Milky Way exist in a rather small range of total mass and mass ratio \citep{Lattimer11,Tauris17}, and so one might naively expect them to drive similar KNe, which therefore ought to be reflected in the SGRB KN population if it is comprised entirely of BNS mergers. Alternatively, it might be that fine differences in the merging binaries or viewer orientation yields strongly differing observables, or that many SGRBs arise from NSBH rather than BNS mergers.

In this paper, we compare the optical and near-infrared light curves of the KN in GW~170817 to a sample of nearby ($z \lesssim 0.5$) SGRBs to ascertain whether or not a KN of similar magnitude to \KN{} could (or even should) have been detected. We use a cosmology of \cosmo{} throughout. All reported errors are $1\sigma$, and given upper limits are $3\sigma$.

\section{\KN{} - The kilonova associated with GW~170817} \label{sec:data}
Our data for \KN{} come from \citet{Tanvir17}. We convert them to absolute magnitudes using a redshift of $z = 0.009783$ \citep{Hjorth17,Levan17}. 

\begin{figure}
\includegraphics[width=8.9cm]{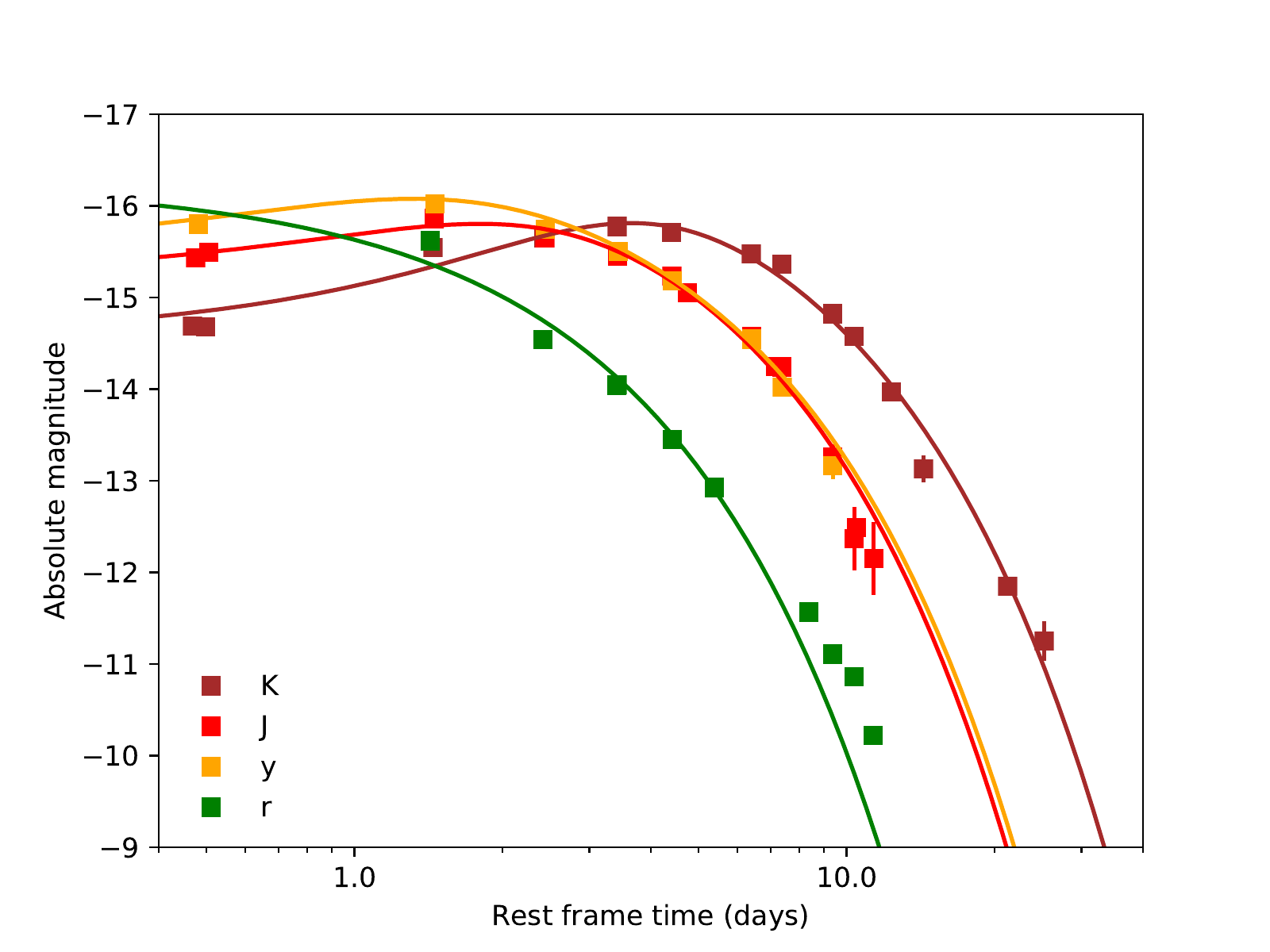}
\caption{Model fit to \KN. The y, J and K-bands are fitted with a Bazin function, while the r-band is better fitted with a simple exponential. Error bars are smaller than the plot symbols in most cases. \label{fig:fit}}
\end{figure}

To represent the evolution of \KN, we fit the light curves with Bazin functions \citep{Bazin11}, which provide an estimate of rise ($\tau_{\rm rise}$) and decay ($\tau_{\rm fall}$) times by fitting the analytical function for a given band
\begin{equation}\label{eq:Bazin}
f(t) = A {e^{-(t-t_0))/\tau_{\rm fall}} \over 1 + e^{-(t-t_0)/\tau_{\rm rise}}},
\end{equation}
where $A$ is the normalisation. \citet{Bazin11} also present an additive constant, $c$, which we find to be zero in each case. We therefore re-fit with the constant excluded. Our fits are shown in Figure~\ref{fig:fit}, while the fit parameters are given in Table~\ref{table:fitparam}.

\begin{table}
\begin{center}
\begin{tabular}{ccccc}
\hline\hline
Band & $A$ & $t_0$ & $\tau_{\rm rise}$ & $\tau_{\rm fall}$ \\
 & $\mu$Jy & h & h & h \\
\hline
r & 544.50 & | & | & 42.27 \\
y & 752.86 & 11.52 & 14.05 & 74.17 \\
J & 652.96 & 24.87 & 20.00 & 71.01 \\
K & 635.33 & 58.20 & 27.21 & 109.87 \\
\hline\hline
\end{tabular}
\caption{The best fits for \KN, taken in the observer frame. The y, J and K-bands are fitted with a Bazin function (Equation~\ref{eq:Bazin}), while the r-band was better fitted with a simple exponential. \label{table:fitparam}}
\end{center}
\end{table}

These fits demonstrate the strongly chromatic behaviour of \KN. Indeed, the r-band shows no apparent peak within the span of our data, and declines from $\sim 1.5$ days as a simple exponential decay. This decay is very different to the power-law seen in GRB afterglows, and coupled with the non-detection of X-rays at comparable times rules out an afterglow component brighter than an r-band absolute magnitude of $M_r \sim -13$ at $0.62$ days, assuming the mean r-band to X-ray flux ratio of 1130 found for SGRBs in \citet{Nysewander09}. Instead, the optical light is probably dominated by a rapidly fading transient created by the synthesis of lanthanide free ejecta, with relatively low opacity \citep{Evans17}. The redder bands are not well explained by such a simple model, and indeed the light curves appear progressively broader as the observations move redward, and can be explained by a lanthanide-rich, higher opacity component \citep{Tanvir17}.

\begin{table*}
\begin{center}
\begin{tabular}{lccclcc}
\hline \hline
GRB & $z$ & References &\vline& GRB & $z$ & References \\
\hline
050509B & 0.2248 & 1, 2, 3 &\vline& 090515 & 0.403 & 28, 29, 30, 31, 32, 33 \\
050709 & 0.161 & 4, 5, 6, 7 &\vline& 100206A & 0.407 & 26, 34, 35, 36, 37, 38, 39 \\
050724 & 0.257 & 8, 9 &\vline& 100625A & 0.452 & 26, 40, 41, 42 \\
051210 & 0.114 & 10, 11 &\vline& 130603B & 0.356 & 43, 44, 45, 46 \\
060502B & 0.287 & 12, 13, 14 &\vline& 140903A & 0.351 & 47 \\
060614 & 0.125 & 15 &\vline& 150101B & 0.134 & 48, 49 \\
061006 & 0.438 & 16 &\vline& 150120A & 0.46 & 50, 51 \\
061201 & 0.084 & 17 &\vline& 150424A & 0.3 & 52, 53, 54, 55, 56 \\
061210 & 0.41 & 18, 19, 20 &\vline& 160624A & 0.483 & 57, 58, 59 \\
070724A & 0.457 & 21, 22, 23 &\vline& 160821B & 0.16 & 60, 61 \\
071227 & 0.384 & 16, 24, 25, 26 &\vline& 170428A & 0.454 & 62, 63, 64 \\
080905A & 0.1218 & 26, 27 &\vline& | & | & | \\
\hline \hline
\end{tabular}
\caption{The sample of SGRBs with their redshifts, $z$, and the source of their photometry.\\
(1) - \citet{Castro-Tirado05}; (2) - \citet{Hjorth05b}; (3) - \citet{Bloom06}; (4) - \citet{Fox05}; (5) - \citet{Hjorth05}; (6) - \citet{Covino06}; (7) - \citet{Jin16}; (8) - \citet{Berger05b}; (9) - \citet{Malesani07}; (10) - \citet{Blustin05}; (11) - \citet{Berger05c}; (12) - \citet{Poole06}; (13) - \citet{Price06b}; (14) - \citet{Rumyantsev06}; (15) - \citet{Yang15}; (16) - \citet{D'Avanzo09}; (17) - \citet{Stratta07}; (18) - \citet{Melandri06}; (19) - \citet{Mirabal06b}; (20) - \citet{Cenko06b}; (21) - \citet{Cenko07}; (22) - \citet{Berger09}; (23) - \citet{Kocevski10}; (24) - \citet{D'Avanzo07b}; (25) - \citet{Berger07b}; (26) - \citet{NicuesaGuelbenzu12b}; (27) - \citet{Rowlinson10b}; (28) - \citet{Morgan09}; (29) - \citet{Updike09c}; (30) - \citet{Cucchiara09c}; (31) - \citet{Siegel09}; (32) - \citet{McLeod09}; (33) - \citet{Perley09c}; (34) - \citet{Leloudas10}; (35) - \citet{Kuroda10}; (36) - \citet{Marshall10}; (37) - \citet{Berger10b}; (38) - \citet{Andreev10}; (39) - \citet{Rumyantsev10}; (40) - \citet{Naito10}; (41) - \citet{Landsman10}; (42) - \citet{Fong13}; (43) - \citet{Tanvir13}; (44) - \citet{Cucchiara13b}; (45) - \citet{deUgartePostigo14}; (46) - \citet{Fong14}; (47) - \citet{Troja16}; (48) - \citet{D'Avanzo15}; (49) - \citet{Fong16}; (50) - \citet{Chornock15}; (51) - \citet{Chester15}; (52) - \citet{Marshall15}; (53) - \citet{Malesani15}; (54) - \citet{Kann15b}; (55) - \citet{Butler15}; (56) - \citet{Knust17}; (57) - \citet{Kuroda16}; (58) - \citet{Kong16}; (59) - \citet{dePasquale16}; (60) - \citet{Jin17}; (61) - \citet{Kasliwal17}; (62) - \citet{Kuin17}; (63) - \citet{Bolmer17}; (64) - \citet{Troja17}.
\label{tab:sample}}
\end{center}
\end{table*}

\section{SGRB Data Sample}\label{sec:SGRBs}
We searched for nearby SGRBs with identified redshifts of $z < 0.5$ in published works \citep{Nysewander09,Fong15} and unpublished archives.\footnote{www.mpe.mpg.de/$\sim$jcg/grbgen.html} For our sample, we define `short' GRBs as those with $T_{90} < 2$~s \citep{Kouveliotou93}, where $T_{90}$ is the duration of the middle 90 per cent of the prompt emission fluence. However, we also include the population of `extended emission' SGRBs. These bursts exhibit negligible spectral lags and hard spectra within the initial pulse(s) inside 2~s (in common with the SGRB population at large), followed by a softer, low intensity tail that inflates $T_{90}$ to several tens of seconds \citep{Norris06,Norris10}. Our identified sample is shown in Table~\ref{tab:sample}. Optical and IR photometric data are then collected from the published literature, or from GCN circulars. Any Vega magnitudes are converted to AB, and corrections for Galactic absorption are performed following the maps of \citet{Schlafly11} and using $R_v = 3.1$ for the Milky Way. We do not make any corrections for extinction in the SGRB host galaxies.

In \KN{} it appears that the optical and IR light is dominated by the KN, at least for several days after the merger, and possibly for longer. Any afterglow component is very weak \citep{Evans17}. However, in SGRBs the situation is very different. It is likely that most SGRBs are viewed close to the jet axis \citep[see e.g.][]{Ryan15}, and so show X-ray and optical afterglows. These afterglows probably dominate the light, at least for the first several days after the merger. However, the early, bright blue and UV emission from \KN{} also suggests that such KN may contribute here. Therefore, it is desirable to estimate the likely afterglow contribution in the optical bands at these times. SGRB afterglows are well established as synchrotron phenomena, and so the X-ray data can be fairly simply extrapolated to optical frequencies. This provides a baseline estimate for the expected magnitude of the afterglow, which can be compared to the optical/nIR observations to check for any excesses that may be symptomatic of a KN (see the shaded regions in Figure~\ref{fig:LCs}), although it suffers from limitations due to the large lever-arm between X-ray and optical frequencies, over which the errors associated with the X-ray observations become significant.

\emph{Swift} X-Ray Telescope (XRT) data are obtained for each SGRB from the $1$~keV flux density light curves on the UK \emph{Swift} Science Data Centre (UKSSDC) burst analyser\footnote{www.swift.ac.uk/burst\_analyser/} \citep{Evans07,Evans09}, except for SGRB~150101B, where we take the \emph{Chandra} data from \citet{Fong16} because the \emph{Swift} light curve is dominated by the Active Galactic Nucleus (AGN) in the host galaxy. Additional X-ray data from \emph{Chandra} and/or XMM-\emph{Newton} are added for SGRBs 050709 \citep{Fox05}, 050724 \citep{Grupe06}, 130603B \citep{Fong14} and 140903A \citep{Troja16}. We correct the \emph{Swift} data for absorption using a ratio of [counts-to-flux unabsorbed]/[counts-to-flux observed], which we obtain from the automatically generated late-time photon counting mode spectral fit in the XRT spectrum repository on the UKSSDC website. The flux densities are converted to AB magnitudes and extrapolated to $6260$\AA{} (r-band) for easy comparison with the optical data. The assumed spectral indices of the extrapolation are $\beta = 0.5$, which represents the case in which the synchrotron cooling break is above the X-ray frequency, and $\beta = 1.0$, which implies the synchrotron cooling break is down near optical frequencies. Both of these assume that the electron energies follow a power-law distribution with an index of $p = 2$, which is the theoretically expected value \citep{Sari98}. In nature, measurements of spectral and temporal indices often imply that this distribution is steeper than $p = 2$ \citep[e.g.][]{Curran10}, suggesting that a steeper spectral index would be needed. However, the synchrotron cooling break is typically measured to be somewhere close to X-ray frequencies rather than in the optical part of the spectrum \citep[e.g.][]{Gompertz15}, so flux densities bracketed by our two extrapolations provide a reasonable estimation of where the SGRB afterglow should occur.

We convert all apparent magnitudes to absolute magnitudes, assuming for a given band $k$, $k_{abs} = k_{AB} - 5\log_{10}(d_L/10 \mathrm{ pc}) + 2.5 \log_{10} (1+z)$. Observed times are divided by $(1 + z)$ to put them in the rest frame. We do not correct for the redshift of the central frequency of the photometric filter because this would require assuming a spectral shape that is uncertain in many SGRBs. Instead we interpolate and shift the well sampled lightcurve of \KN\ to the rest frame wavelength represented by the observations (see below). This means our photometry is presented in its observer frame band, such that the magnitudes are given at $\lambda_k = (1+z) \lambda_{r}$, where $\lambda_r$ is the rest frame emitting wavelength.  Since KNe do not appear (or are at least very weak) at X-ray frequencies, this issue does not affect our X-ray extrapolations.

While the optical to IR spectral energy distributions of our SGRBs are usually poorly constrained, we do have good colour coverage of \KN. We therefore adjust our model fits to the wavelengths of the SGRB observations via linear interpolations between the fitted curves in Figure~\ref{fig:fit}. Where the rest-frame frequency is bluewards of the r-band fit to \KN, we supplement with observations with the \emph{Swift}-UVOT u-band data from \citet{Evans17} and the HST F475W (g-band) data from \citet{Tanvir17}, with an early g-band point based on the spectral flux density measured in our MUSE spectroscopy. We fit both datasets with an exponential decay to ascertain the required colour correction. Fitting in the observer frame, we find $A_g = 321.92$~$\mu$Jy, $\tau_{g, fall} = 39.10$ hours, and $A_u = 321.39$~$\mu$Jy, $\tau_{u, fall} = 20.25$ hours. We note that these bands have only a small number (3-4) points to fit, and so we have assumed they follow an exponential decay as the r-band does, but do not have sufficient data discriminate between alternative models. We do not extrapolate beyond the u-band if the rest-frame wavelength is less than 3560\AA.

\begin{figure*}
\includegraphics[width=8.9cm]{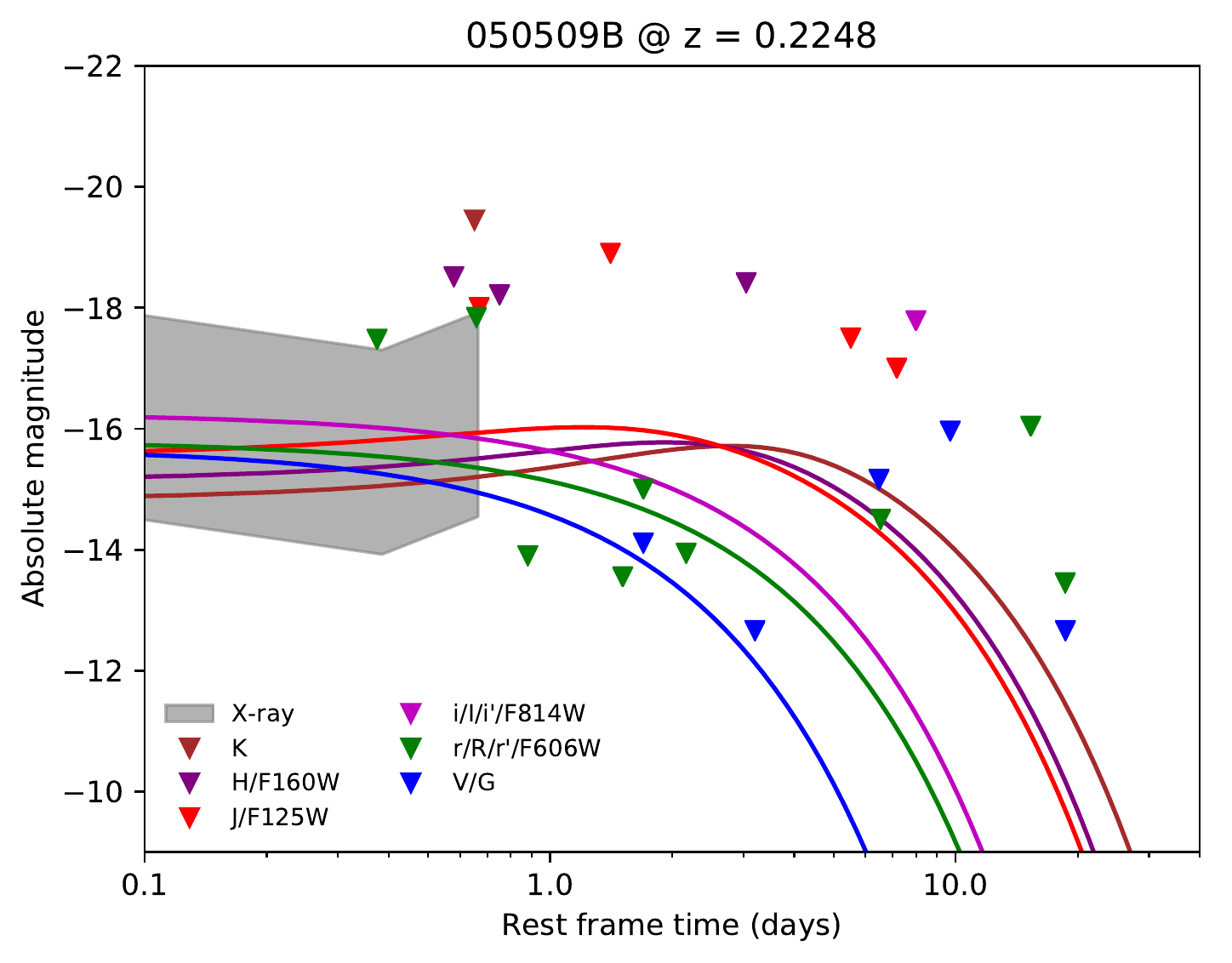}
\includegraphics[width=8.9cm]{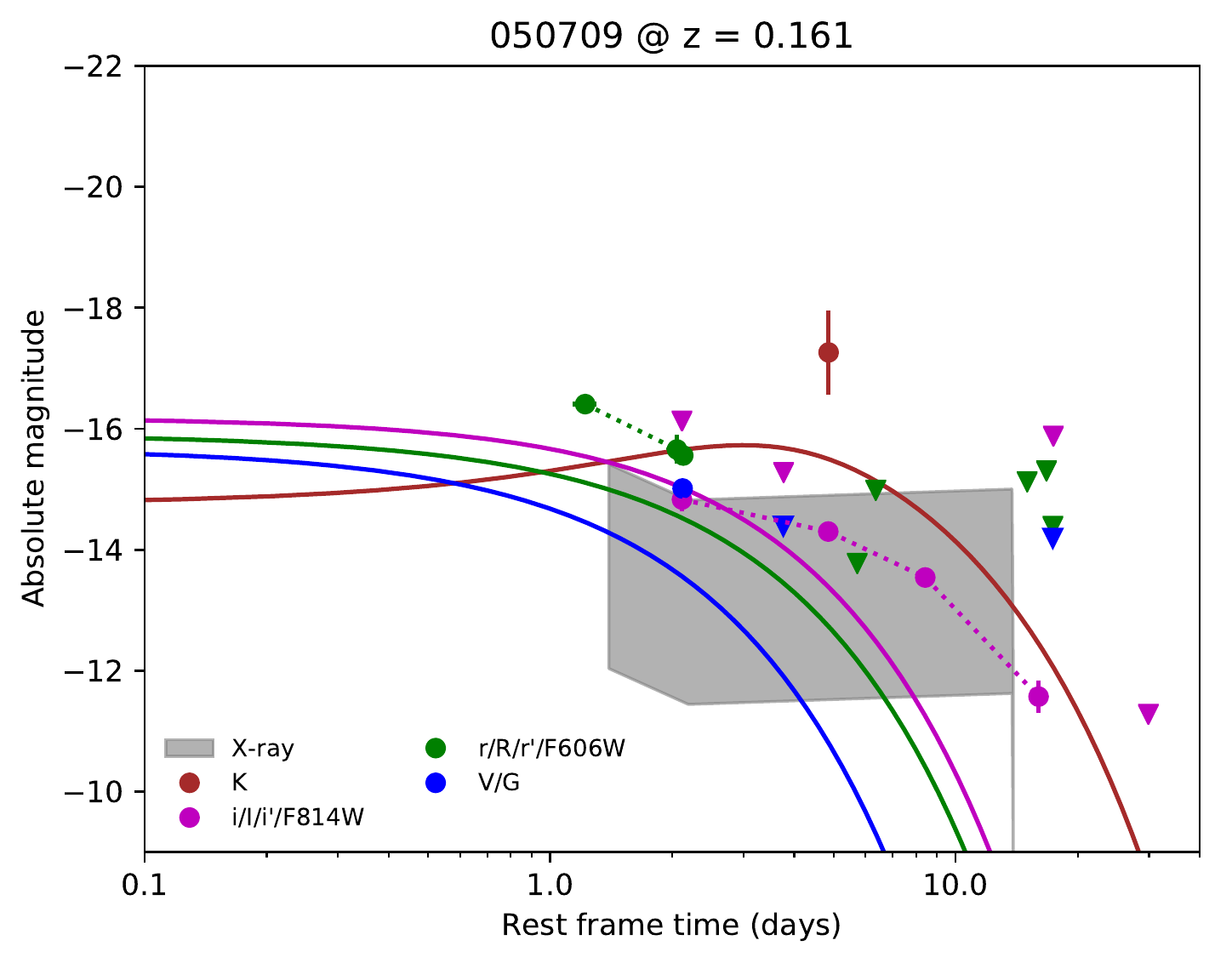}
\includegraphics[width=8.9cm]{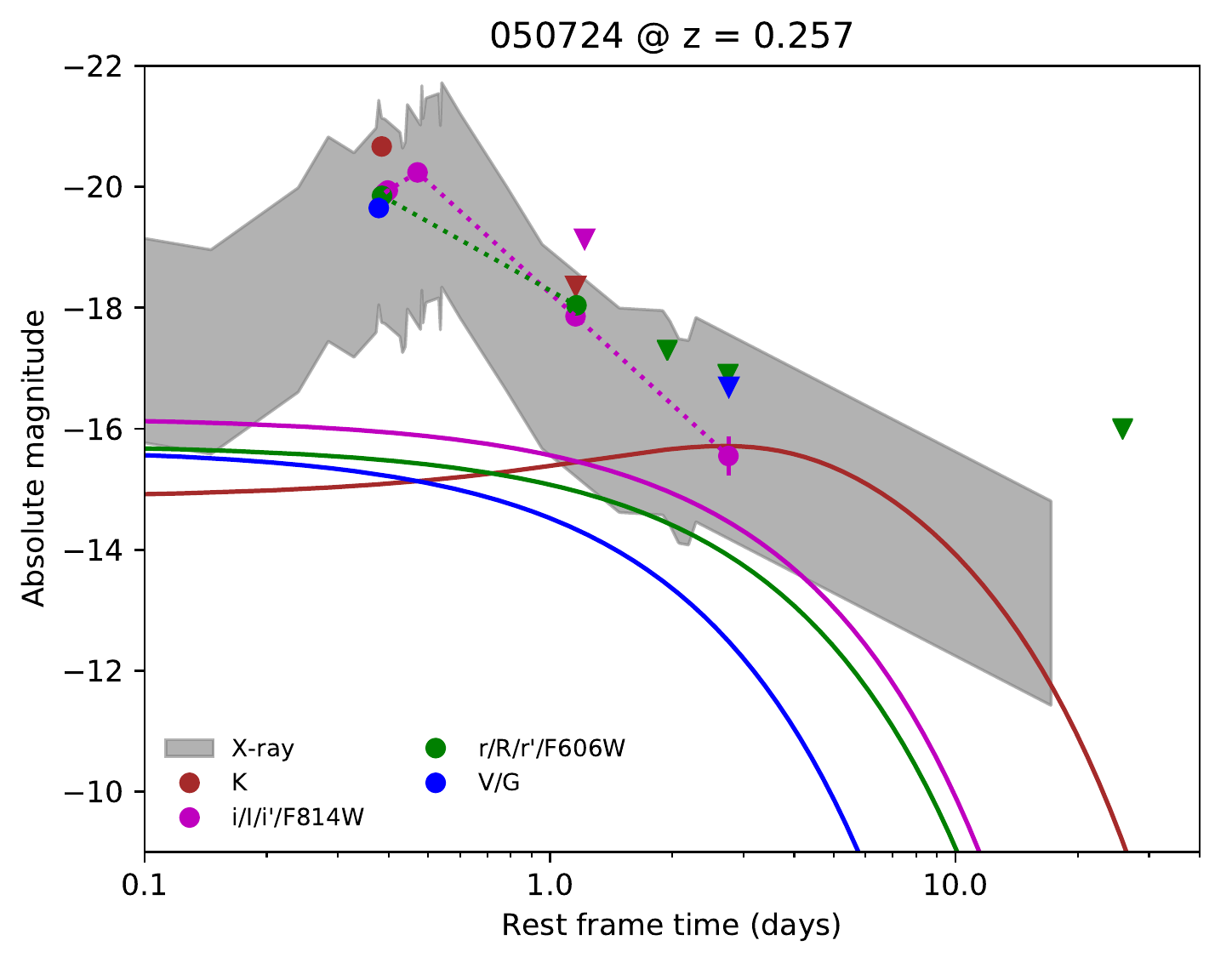}
\includegraphics[width=8.9cm]{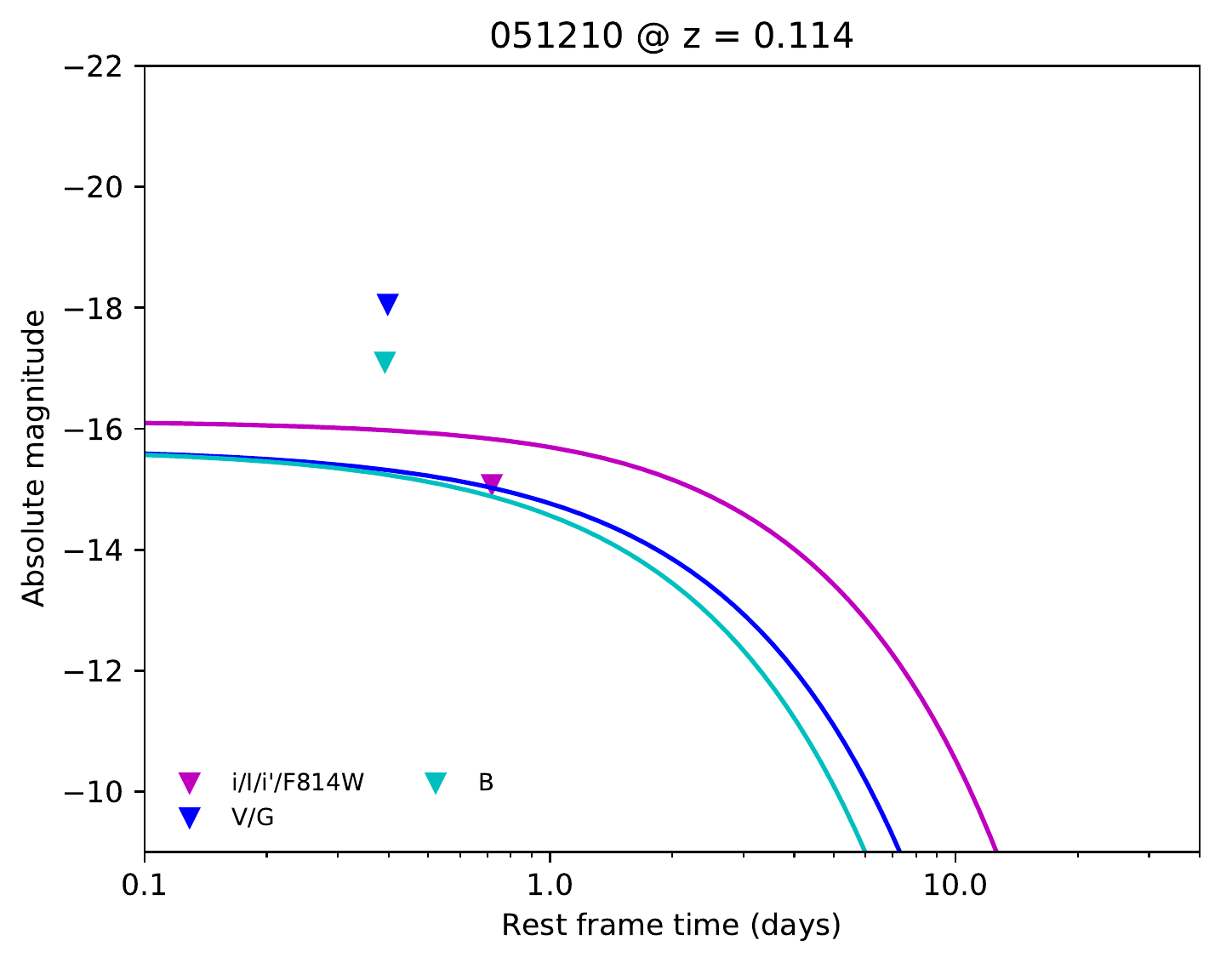}
\includegraphics[width=8.9cm]{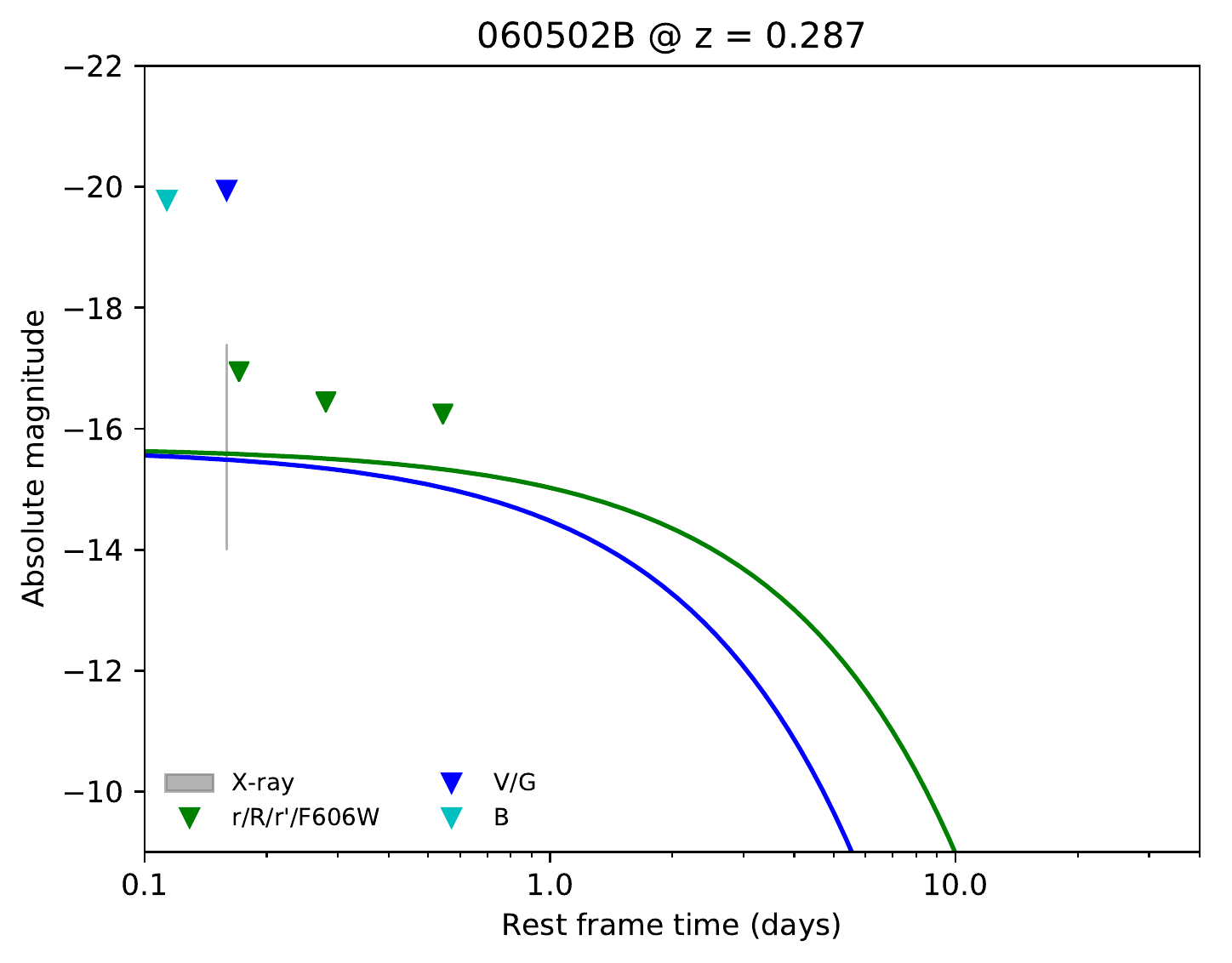}
\includegraphics[width=8.9cm]{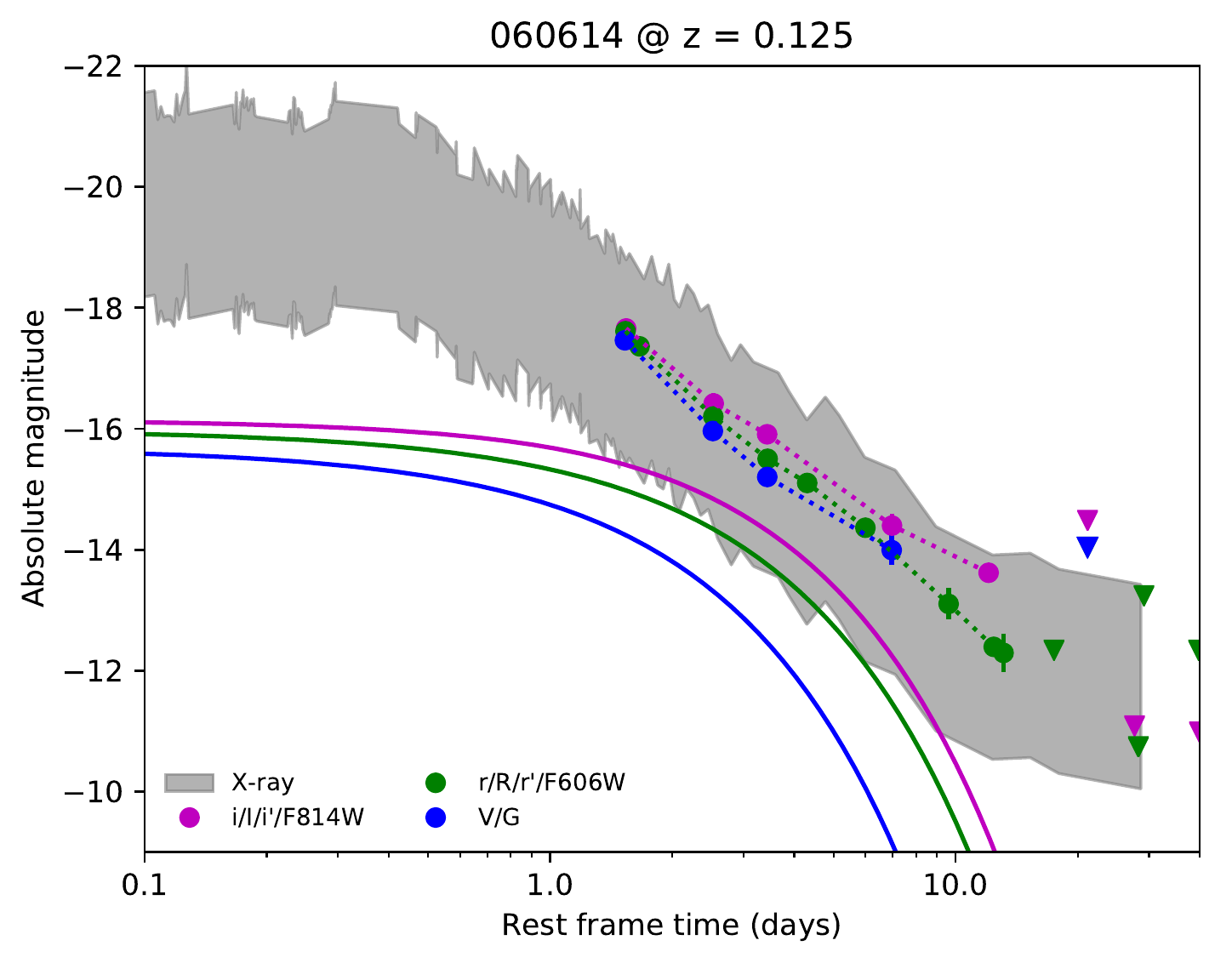}
\caption{Our sample of SGRB afterglows compared to the model fits of \KN. Coloured circles are detections, and triangles are upper limits. These data are presented in their observer frame filters, and the \KN{} models (coloured lines) are shifted in frequency to match them using linear interpolation of the fits in Figure~\ref{fig:fit} supplemented by the g-band data from \citet{Tanvir17} and the u-band data from \citet{Evans17}. We do not extrapolate if the rest-frame wavelength is below the u-band (3560\AA), so some data do not have models plotted. The grey band is the extrapolation of the X-ray flux to 6260\AA{} (r-band). SGRBs from Table~\ref{tab:sample} with no constraining observations are not included. \label{fig:LCs}}
\end{figure*}
\setcounter{figure}{1}
\begin{figure*}
\includegraphics[width=8.9cm]{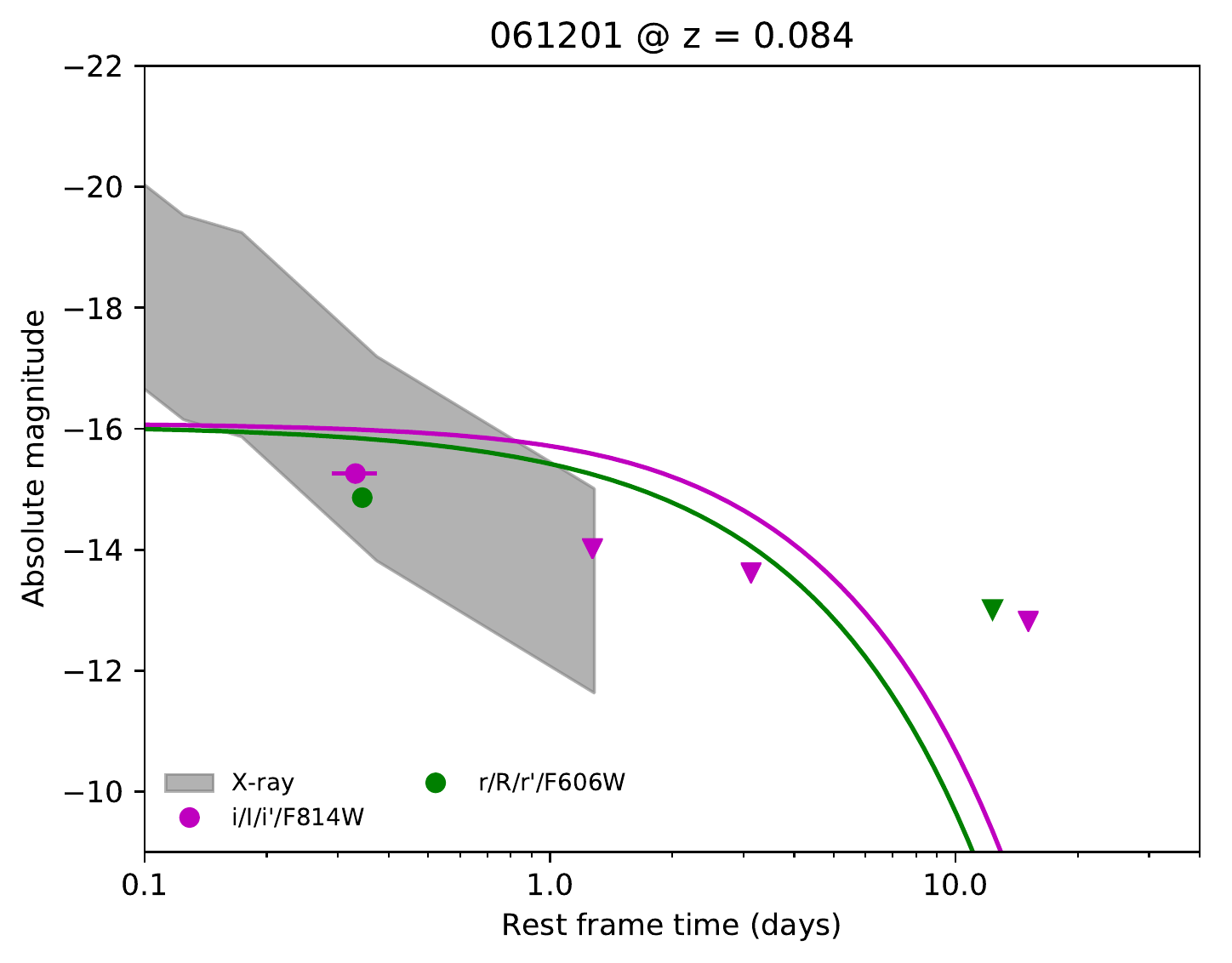}
\includegraphics[width=8.9cm]{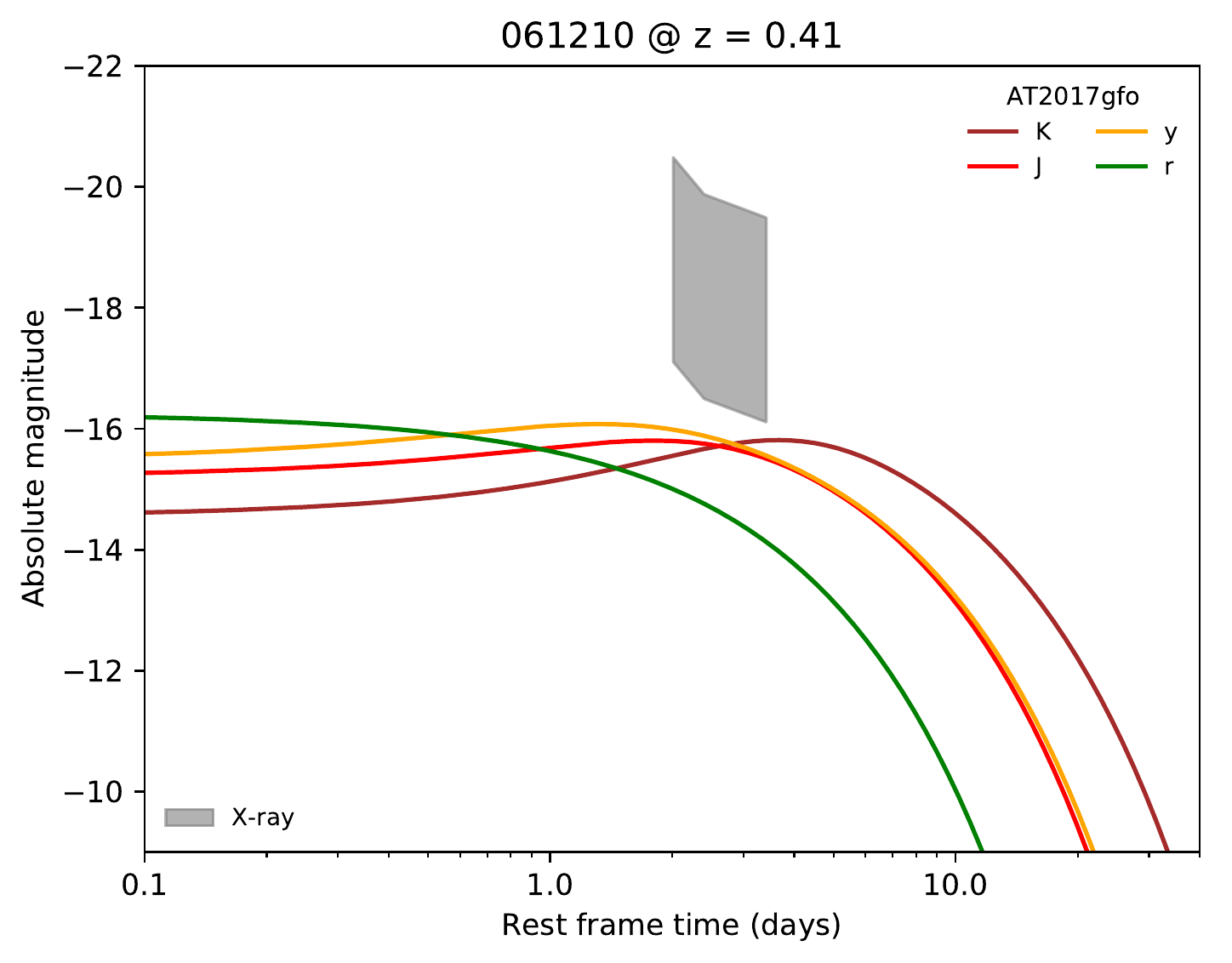}
\includegraphics[width=8.9cm]{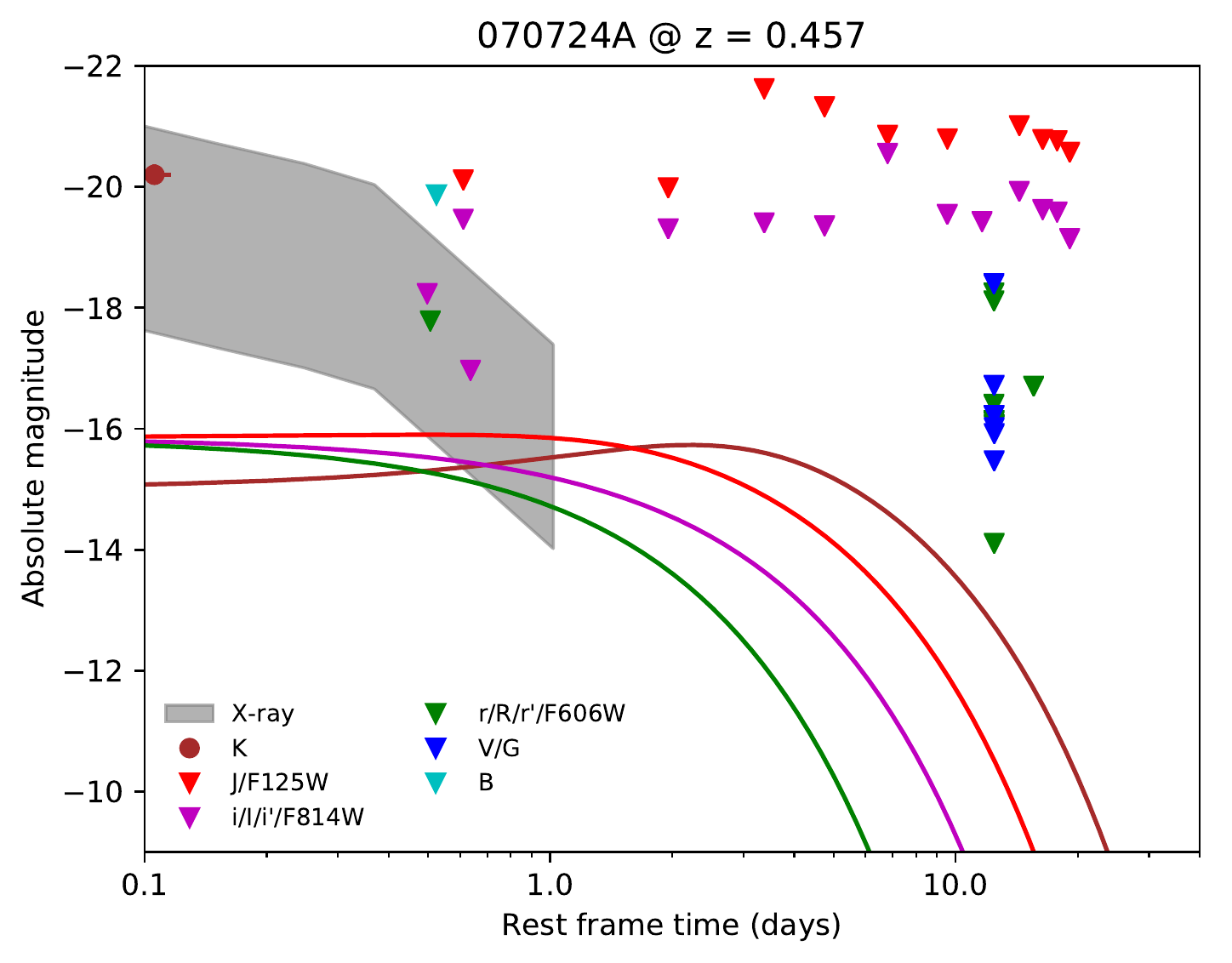}
\includegraphics[width=8.9cm]{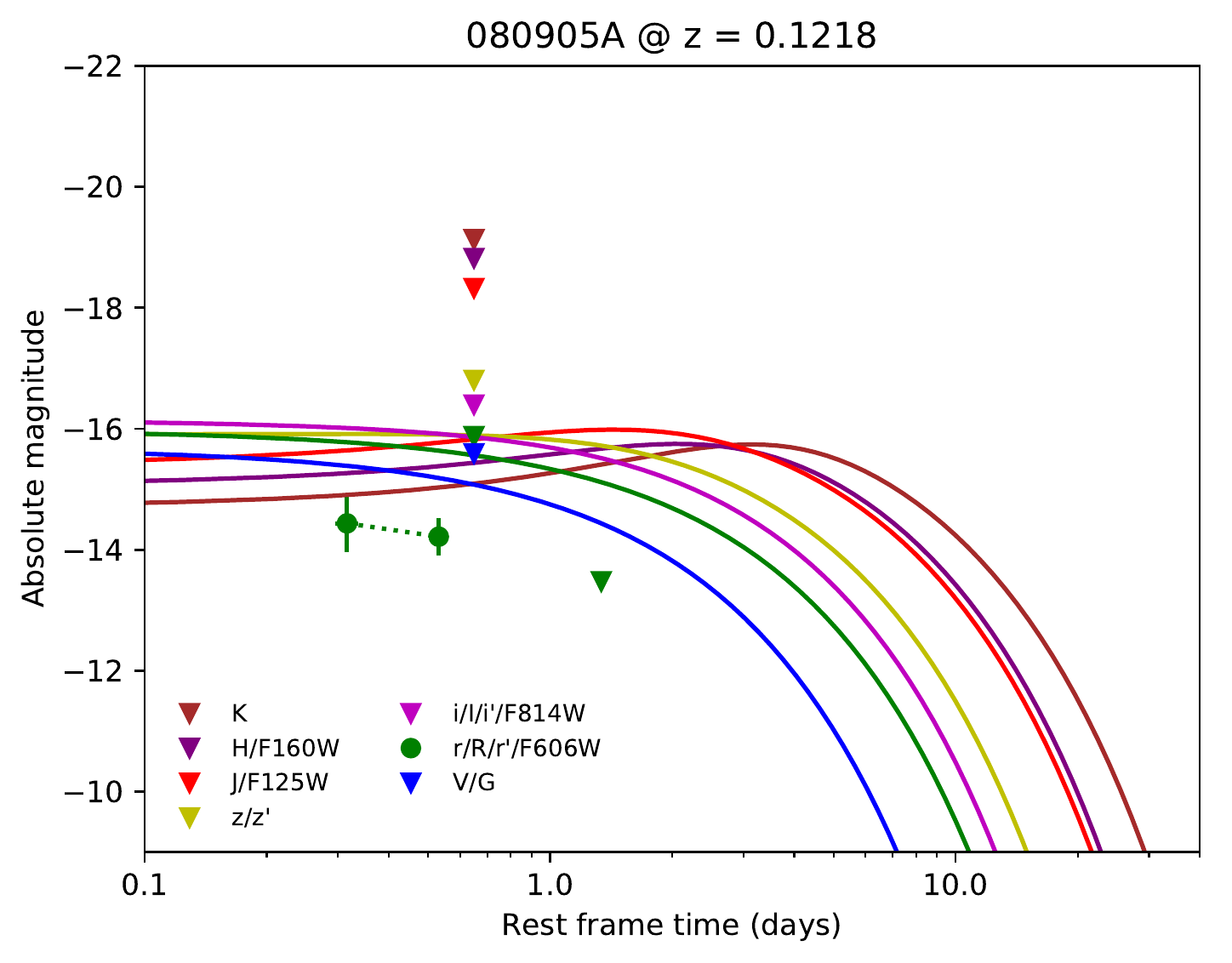}
\includegraphics[width=8.9cm]{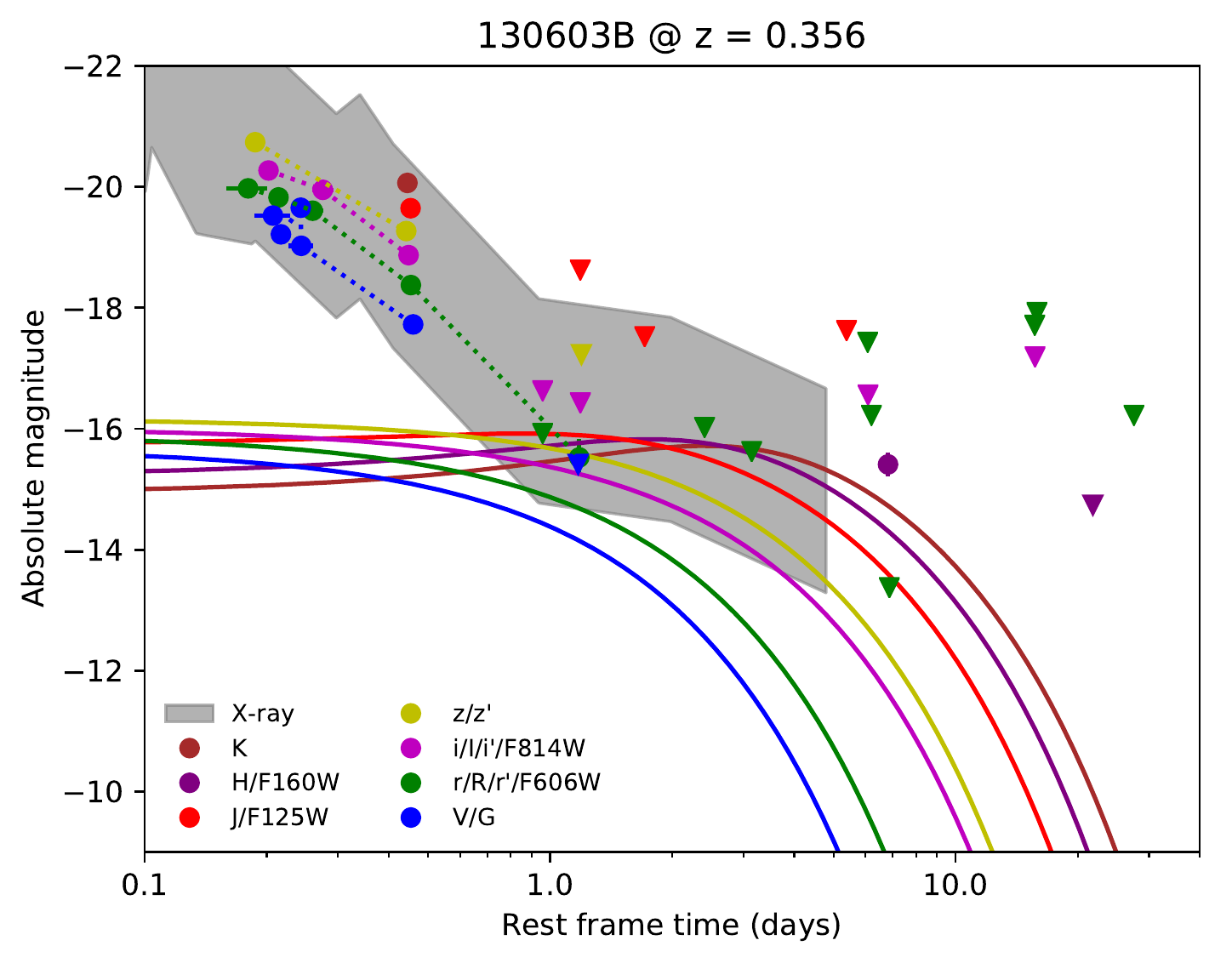}
\includegraphics[width=8.9cm]{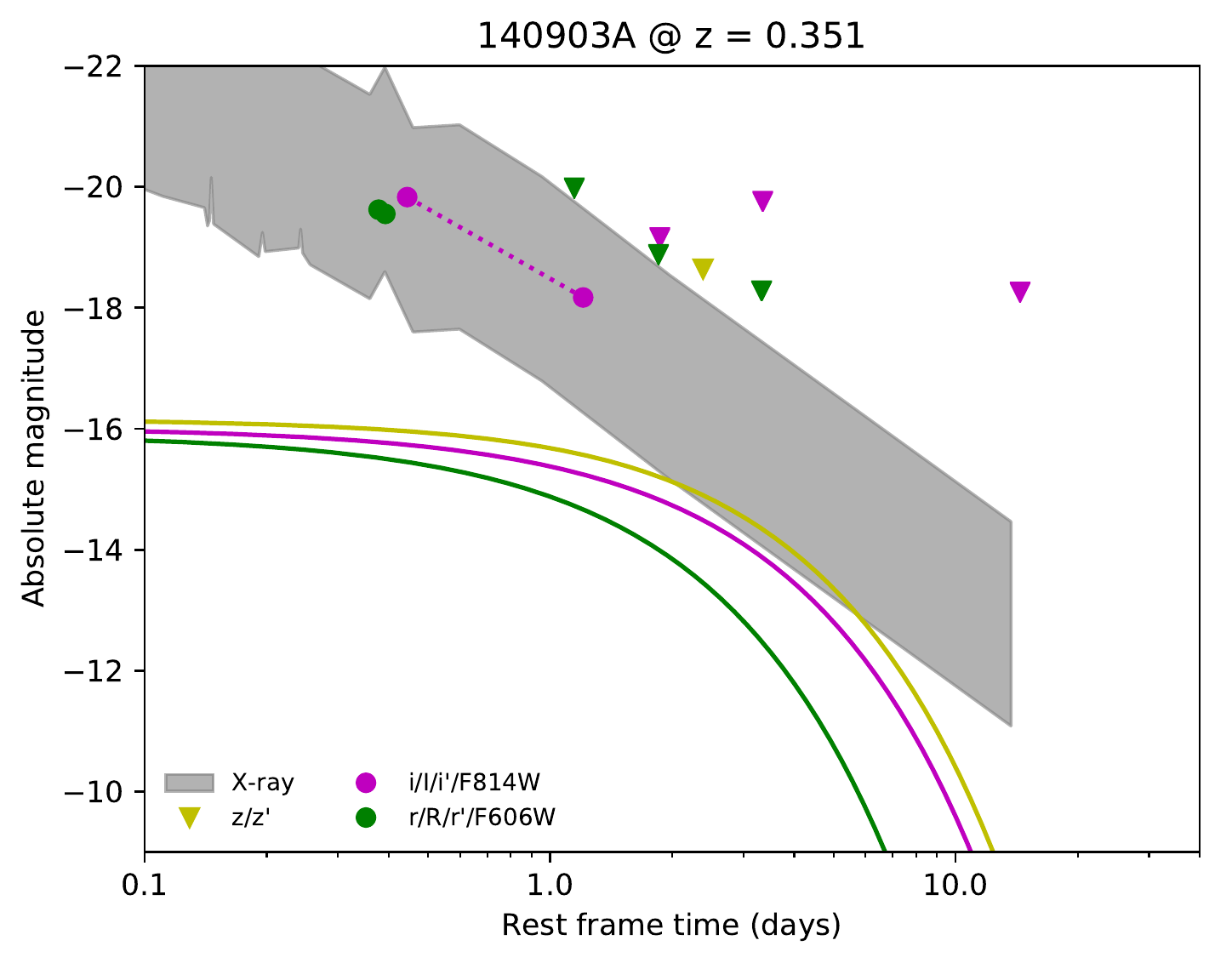}
\caption{continued}
\end{figure*}
\setcounter{figure}{1}
\begin{figure*}
\begin{center}
\includegraphics[width=8.9cm]{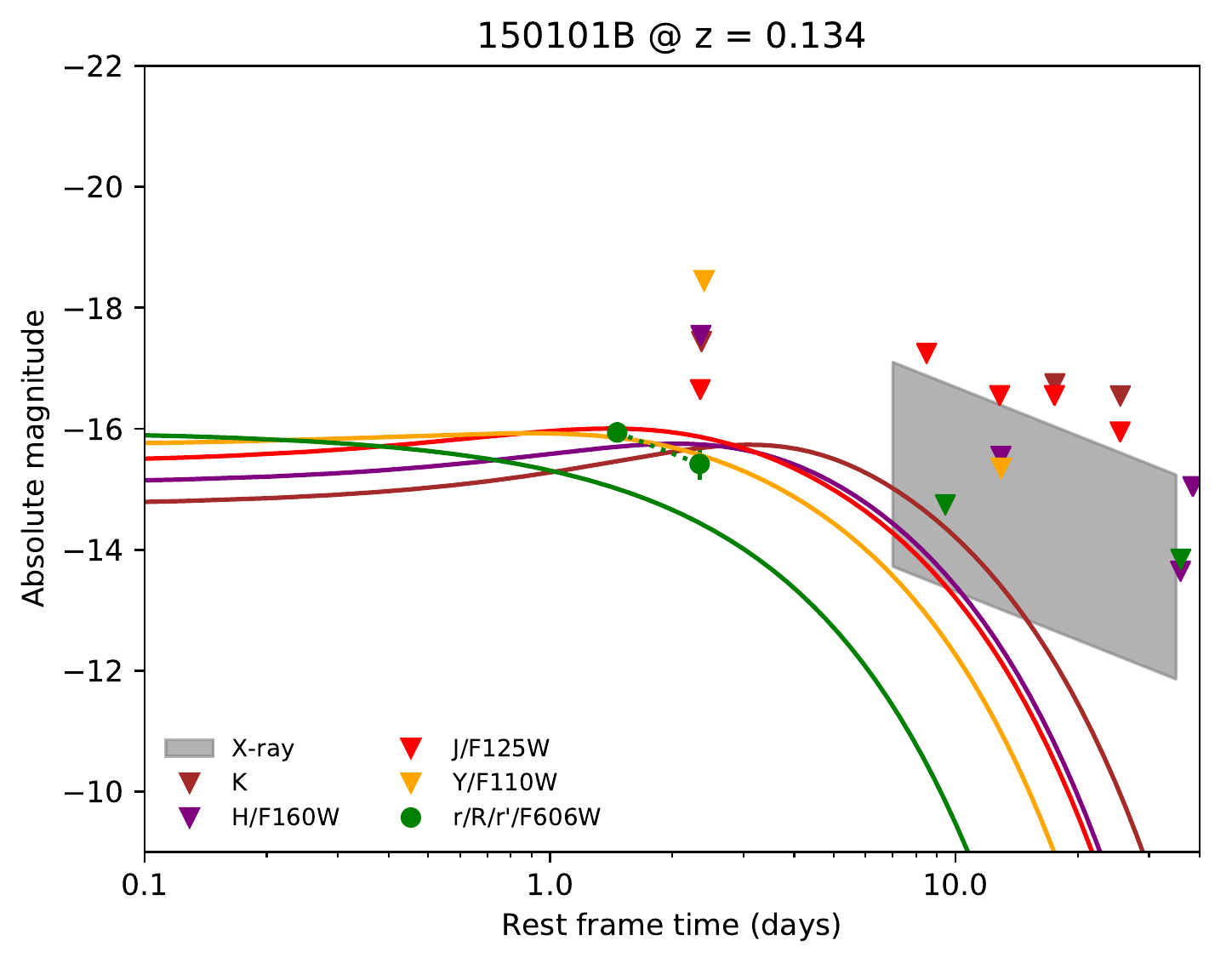}
\includegraphics[width=8.9cm]{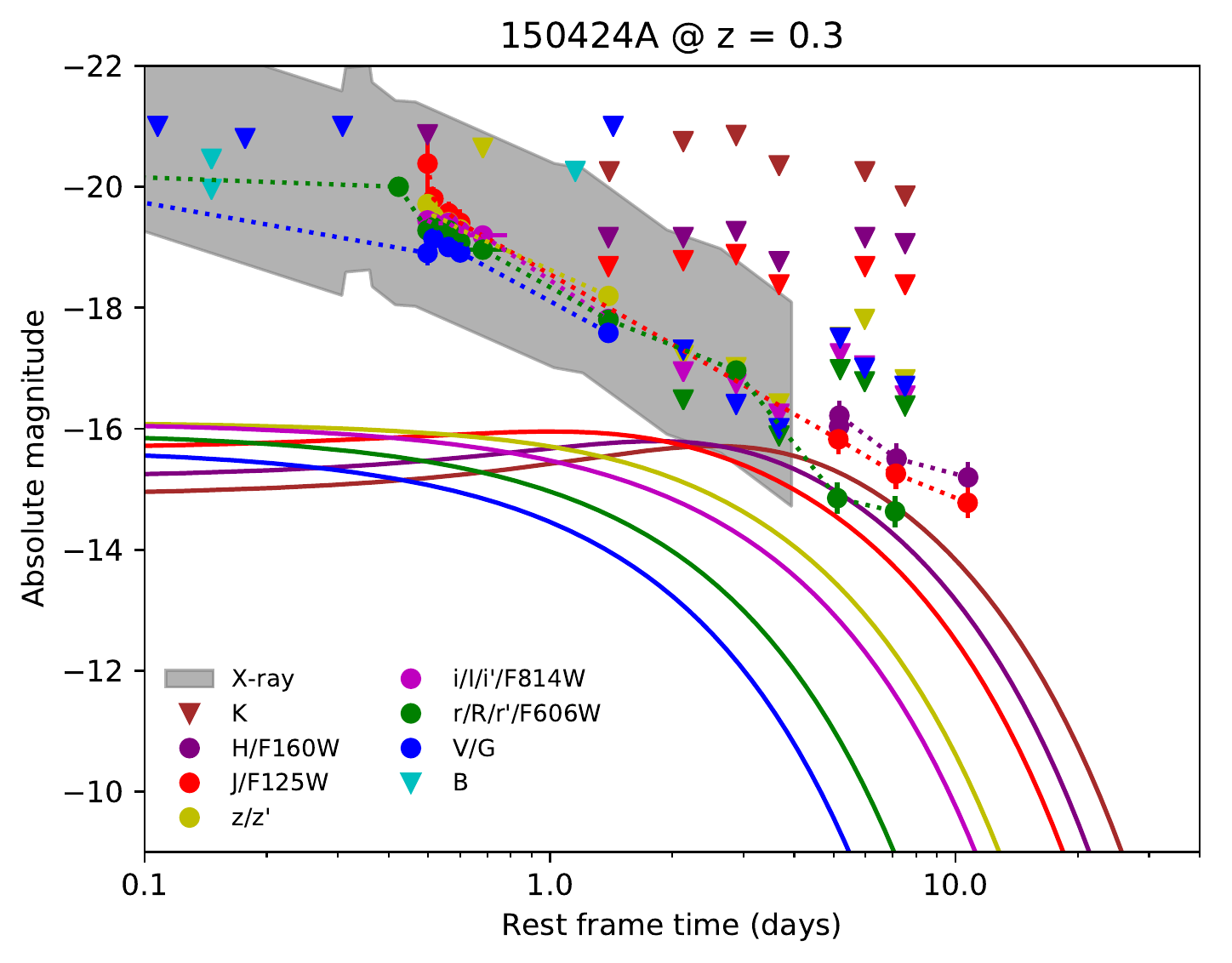}
\caption{continued}
\end{center}
\end{figure*}

\section{Results}
Light curves of our results are shown in Figure~\ref{fig:LCs}. Based on these light curves, our sample broadly divides into four main categories:
\begin{enumerate}
\item{SGRBs with deep limits constraining to an \KN-like KN.}
\item{SGRBs with candidate KNe.}
\item{SGRBs with afterglow detections bright enough to mask an \KN-like KN.}
\item{SGRBs with no constraining observations.}
\end{enumerate}

\subsection{SGRBs with deeper limits than \KN}
We find that for several SGRBs, a KN as bright as \KN{} could have been detected. For four bursts in particular, deep $3\sigma$ upper limits that a factor of two or more fainter than the detections of \KN{} at comparable rest-frame times appear to rule out a KN like \KN. They are SGRBs 050509B (3 times fainter than AT2017gfo), 051210 (2 times fainter), 061201 (4 times fainter) and 080905A (4.5 times fainter).


In each case, a KN similar to \KN{} could have been detected if it had been present. These limits assume that in each case the redshift of the burst is correctly ascribed, and this is considered in more detail in Section 5. It should also be noted again that no host galaxy extinction has been included in our study.

\subsection{SGRBs with candidate KNe}
Compared to the SGRB KN candidates, \KN{} appears to be faint, though the contribution of the SGRB afterglow to the observed flux in the bursts is uncertain. The H-band detection in GRB 130603B is almost $3$ times brighter than the interpolated KN model fit at the time of the observation. Similarly in SGRB 050709, the K-band detection is far brighter than the K-band of \KN. However, the i-band light curve appears to be fainter and peak later, though the photometry is limited. The afterglow of GRB 060614 is already much brighter than the light curve of \KN, and the i-band photometric excess \citep{Yang15} is over $50$ times brighter than the contemporary KN fit, though it could be a later/longer rise like SGRB 050709. Alternatively, as shown in Figure~\ref{fig:LCs}, the extrapolation of the X-ray afterglow to optical/IR wavelengths may imply significant afterglow contributions at this time, such that the KN components can be fainter. The photometric excess claimed by \citet{Jin17} in 160821B is the only potential KN less bright than \KN{} \citep[see][]{Troja16b}, though the data are inconclusive in this case (see Tanvir et al. in prep. for a detailed analysis of this burst).

\subsection{SGRBs with bright afterglows}
We identify several SGRBs in which the afterglow was likely too bright for a \KN-like KN to be detected. SGRB~150424A has optical/nIR afterglow detections several magnitudes brighter than \KN{} \citep[see][]{Tanvir15}, and its detections and limits are never less than about $3$ times brighter than the relevant model light curve in the H-band, or $10$ times in the r-band. This burst may also be at a higher redshift than reported here \citep{Tanvir15}, which would mean the afterglow is even brighter when corrected for distance. SGRBs 140903A and 150101B both have extrapolated X-ray afterglows that are brighter than \KN, with optical detections that support the extrapolation, though there are no nIR limits constraining a redder transient. The i-band detection in SGRB~140903A is almost $15$ times brighter than the \KN{} model. The r-band detection in SGRB~150101B is just $2$ times the flux of the \KN{} model, so could potentially have a KN contribution. It is notable that its decay between the two epochs of observations is very simiar to \KN. SGRB~050724 features a large flare seen in X-ray, nIR and optical bands. The i-band detection close to $3$ days is of a similar brightness to the one in 060614 \citep[which was identified as a KN by its i-band excess;][]{Yang15}, but the contribution from the various emission features is unknown.

\subsection{SGRBs with no constraining observations}
In the final category, six SGRBs do not have sufficiently deep observations to place any meaningful constraints on the presence of a KN like \KN.

SGRB~061210 may potentially have a bright afterglow if the extrapolated X-ray flux is in fact representative, but no observations in the optical/nIR are available. Neither SGRB 070724A nor 060502B have optical or nIR limits deep enough to provide meaningful constraints on the \KN{} models. However, the r-band limits in 060502B are within a factor of $2$ of the r-band model of \KN, suggesting that any KN in this burst is not significantly brighter than \KN. SGRBs 061006, 071227 and 170428A all lie on bright host galaxies. No image subtraction of the host galaxy has been performed, and so any contributions from the afterglow or a possible KN are swamped by the light of the host.

\subsection{Colours}
One of the more distinct features of \KN{} was its colour evolution. The optical transient began blue, and slowly became redder over the course of several days \citep[see e.g. Figure 3 from][]{Tanvir17}. This was explained as a two (or possibly three) component KN: a `blue' component from high velocity lanthanide-poor dynamical ejecta from the poles \citep[e.g.][]{Nicholl17}, and a slower `red' lanthanide-rich component driven in an isotropic wind, and rising to prominence later. Some groups also invoke an intermediate `purple' component \citep[e.g.][]{Villar17}.

Many of the SGRBs in our sample feature multi-colour detections, and this allows us to compare their colours to \KN. Seven feature photometric detections in more than one filter that are approximately contemporaneous. They are SGRBs 050709, 050724, 060614, 061201, 130603B, 140903A and 150424A. We compare their colours to our model curves of \KN. Colours are compared in the rest frame, and so are k-corrected from their initial redshift.

From the KN candidates, SGRB 050709 appears to be bluer than \KN{} at around 2 rest-frame days after trigger when measured in g-r ($0.55 \pm 0.12$ in the SGRB data versus $\sim 0.96$ in our \KN{} model) and r-i ($-0.30 \pm 0.22$ versus $\sim 0.50$). However, at around 5 days after trigger its I-K colour of $2.96 \pm 0.70$ is comparable within errors to our model fits, where I-K $\sim 2.12$. Whether this is due to colour evolution is uncertain because we have no K band magnitude measurements from the earlier epoch. At 2 days, SGRB 050709 is brighter in r than in either g or i. This is unusual for an SGRB afterglow, and potentially suggests some spectral evolution. SGRB 060614 is also bluer than \KN{} early on, with g-r $= 0.15 \pm 0.04$ in the data compared to $\sim 0.74$ in the model, and r-i $= 0.05 \pm 0.04$ in the data versus $\sim 0.42$ in the model at around $1.5$ rest-frame days. It shows a redwards linear evolution in both colours in a similar way to \KN, becoming g-r $= 0.30 \pm 0.10$ ($\sim 1.31$) and r-i $= 0.41 \pm 0.11$ ($\sim 0.56$) over the next two days. By 2 weeks after trigger, the colours have converged, with r-i $=1.23 \pm 0.18$ in the data and r-i $\sim 1.11$ in the model. This may indicate that the KN in 060614 now dominates the observed emission. Multi-wavelength coverage for SGRB 130603B is only available inside one day after trigger. This emission has already been shown to be consistent with an afterglow by \citet{Fong14}. The r-i colour is $0.45 \pm 0.06$, compared to $\sim 0.19$ for the contemporaneous \KN{} model at $0.2$ days.

At almost half a day after trigger, the r-i colour of the model is $\sim 0.25$, and three of the remaining four SGRBs with colour information are broadly consistent with this: 061201 has r-i $0.40 \pm 0.14$; 140903A has r-i $= 0.28 \pm 0.08$; and 150424A has r-i $= 0.16 \pm 0.14$. The exception, 050724, exhibits a large flare at this time (Figure~\ref{fig:LCs}). There is no evidence for significant evolution beyond this time, in most cases due to a lack of continuing multi-colour monitoring. SGRB 150424A has r-i $= 0.01 \pm 0.15$ at $1.4$ days, consistent with no evolution, while the \KN{} model has evolved to r-i $\sim 0.65$ at this time. \citet{Jin17} also found this SGRB to be consistent with no chromatic evolution.

\section{Discussion}
We search our sample of 23 SGRBs with identified redshift of $z < 0.5$ and optical/nIR observations for detections and limits constraining to KNe similar to \KN{}, or bright afterglows capable of masking a KN of this magnitude. 3 have claimed KNe in the literature (050709, 060614, 130603B) with a further marginal case (160821B; see Tanvir et al. in prep. for a detailed analysis). 4 have limits deeper than \KN{} (050509B, 051210, 061201, 080905A), and 3 of these are over a magnitude deeper (050509B, 061201, 080905A). 1 more has limits of comparable depth (060502B). 4 have bright optical/nIR detections consistent with light arising from an afterglow component (050724, 140903A, 150101B, 150424A), although at least in one case at magnitudes only marginally brighter than \KN\ (GRB 150101B). 2 have afterglows that are implied to be bright by the X-ray extrapolation (061210, 070724A), although were in fact not observed \citep[in the case of 070724A, the absence of the afterglow has been suggested to be due to the presence of dust;][]{Berger09}. 3 have bright host galaxies (061006, 071227, 170428A), and the remaining 5 are completely unconstrained by the available observations. The broad range of magnitudes in the sample suggests a diversity in the brightness of KNe associated with SGRBs.

For three bursts in particular, the absence of a KN is conspicuous. SGRBs 050509B, 061201 and 080905A all have limits much deeper than the detections for \KN, and any KN would have had to have been at least five times fainter to be missed, yet \KN{} itself is fainter than any KN seen in SGRBs so far. Our findings for SGRB 050509B are consistent with \citet{Fong17}.

There are two concerns that should be addressed in interpreting these limits. The first is whether there could be some unseen extinction along the line of sight. In the case of SGRB 050509B and 051210 this is of particular concern because there is no optical light at any epoch, and too few X-ray photons to determine a column density. However, SGRB 050509B is spatially coincident with the outskirts of a giant elliptical galaxy \citep{Castro-Tirado05,Gehrels05,Hjorth05,Bloom06}, which is the putative host (and the source of the redshift). Given its location, it is unlikely that dust extinction plays a significant role. The apparent faintness of the optical/nIR emission is therefore more likely to be intrinsic, perhaps due to a sparse local environment or a wide viewing angle from the SGRB jet. For SGRB 061201 and 080905A the detection of the sources in optical light rules out extreme extinction, but does not discount moderate levels. \citet{Fong15} find that both of these bursts are consistent with having no host extinction in their modelling, and this suggests that the limits on a KN are indeed as constraining as they appear to be. It is notable in these examples that the optical light is consistent with the X-ray extrapolation (where available) with $\beta = 0.5$, and so the bursts are not ``dark'' GRBs \citep{Jakobsson04}. 

The second concern is the validity of the assumed redshifts. Only one short GRB has a redshift measured in absorption, GRB 130603B \citep{deUgartePostigo14}. In other cases the redshifts are based on putative host galaxy identifications. For GRB 050509B there is a large cD galaxy close to the location, and a coincidence with a massive, merging galaxy cluster. Much of the mass along this line of sight lies within this cluster, and the redshift has high confidence \citep[see e.g.][]{Bloom02,Levan07}. For GRB 080905A the burst position overlaps the spiral arm of the host galaxy, again suggesting a chance alignment probability of $\lesssim 1\%$. However, for GRB 061201 the situation is more complex. The burst belongs to the so-called hostless SGRB population \citep{Berger10,Tunnicliffe14}, and so the redshift is based on a proximity to the Abell 995 cluster \citep{Berger06,Stratta07}. This is a rich cluster, but there are no galaxies within a few arcseconds of the burst position, and so the probability of chance a alignment is significant, and the redshift should be viewed with caution. While the probabilities of any given burst redshift being wrong appear small, it should also be noted that assigning redshifts by probabilistic arguments favours the brightest nearby galaxies, many of which are likely to be closer, and so this may produce a bias in which incorrect host assignments, lead to lower redshifts, and hence strong KN constraints. 

The SGRBs with bright afterglows still may have KN contributions in the detected flux. The inferred contribution from an \KN-like KN ranges from $1/15$ (140903A, i-band) to $1/2.5$ (150101B, r-band), and the latter end of the scale is certainly enough to cause a spectral energy distribution (SED) to deviate away from the power law expected from the synchrotron afterglow. A KN was discovered this way for both SGRBs 050709 \citep{Jin16} and 060614 \citep{Yang15}, and potentially 160821B too \citep{Jin17}. However, no multi-colour observations are available for any of these bursts except 150424A, the brightest of the four. \citet{Jin17} found no evidence of chromatic deviation in this source.

\KN{} appears to be somewhat fainter than other established KN, but the lack of X-ray afterglow to deep limits \citep[$2.7\times10^{-13}$~erg~s$^{-1}$~cm$^{-2}$ at $0.62$ days after trigger;][]{Evans17} and constraints from the GW data suggest that the viewing angle is further off-axis than in typical SGRBs - perhaps up to $28$ degrees away from the jet axis \citep{Evans17,Haggard17,Ligo17,Margutti17,Tanvir17}. The fainter KN in GW~170817 is therefore within the variation expected from the observer position \citep[e.g.][]{Grossman14}. The peak time and luminosity of a KN is also a function of ejecta mass and opacity \citep[e.g.][]{Metzger10b}, and naturally some variation is expected from case to case.

\section{Conclusions}
\begin{figure}
\includegraphics[width=8.9cm]{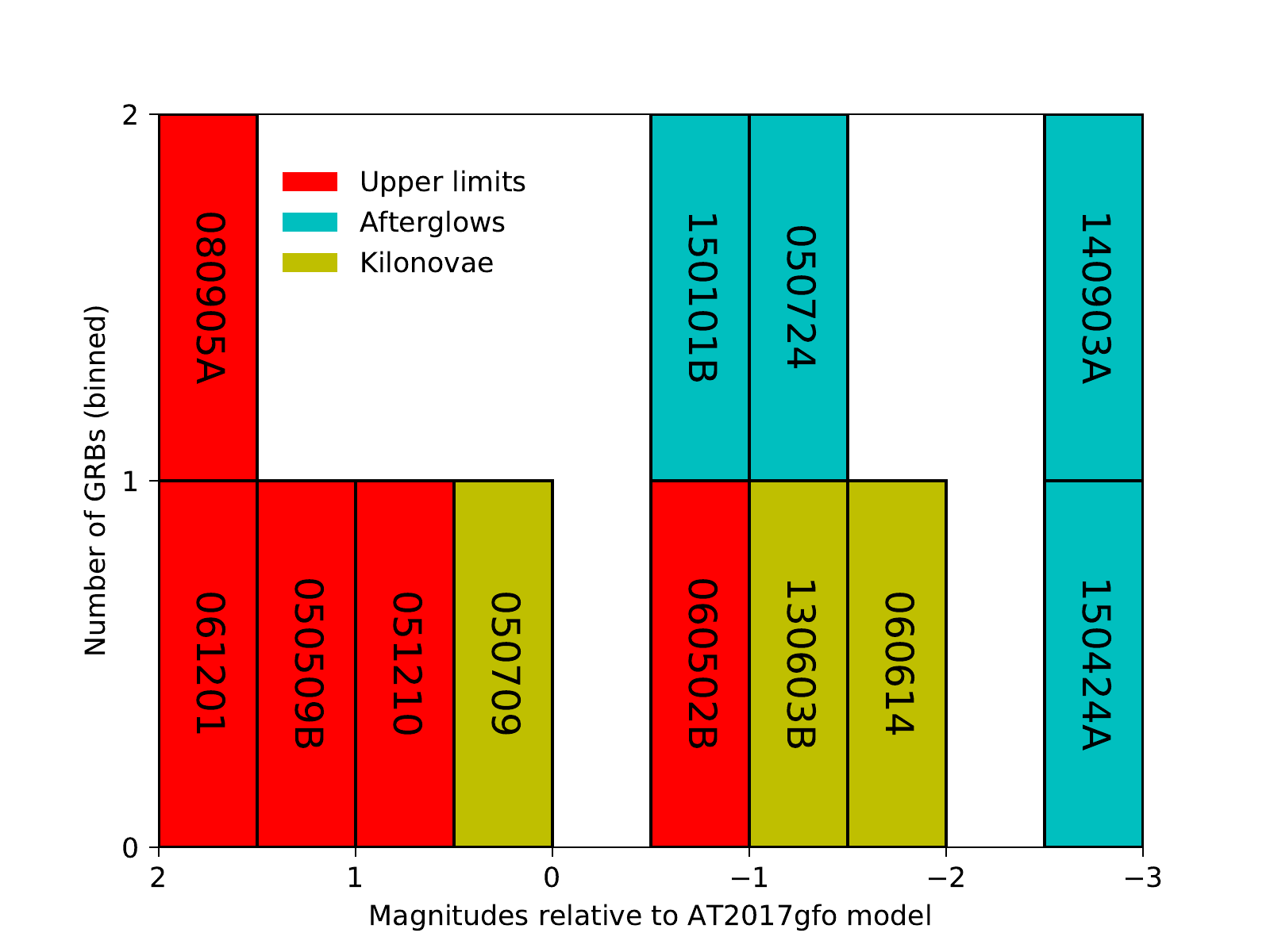}
\caption{A snapshot of the i/r-band magnitudes of our SGRB observations relative to the \KN{} models, taken between $0.5$ and $3$ days after trigger. Data are binned into $0.5$ magnitude intervals. Bursts with data not constraining to the \KN{} models are not included. `Upper limits' (red) refers to bursts with deep limits constraining to an \KN-like KN. `Afterglows' (blue) are the SGRBs with afterglows bright enough to mask \KN. `Kilonovae' (yellow) are the SGRB KN candidates. \label{fig:compare}}
\end{figure}

Our analysis reveals a diverse range of KN possibilities, as in some SGRBs we find upper limits for optical/nIR emission several magnitudes deeper than \KN, in others there are identified (or suspected) KNe that are brighter, and we also find SGRBs with bright afterglows capable of masking KNe that are brighter still. Our sample spans five magnitudes at a few days after trigger (Figure~\ref{fig:compare}). The most interesting comparison is between the SGRBs with detected KNe (including \KN) and the $3$ SGRBs with deep limits; the diversity between these two groups is hard to reconcile with the highly uniform distribution of known BNS masses and mass ratios in the Milky Way \citep{Lattimer11,Tauris17}.

The relatively small range of viewing angles in the SGRB population means that the observer position alone probably can't explain the $\sim 3.5$ magnitudes (a factor of 25 in flux) between the KN in SGRB 060614 and the upper limits in SGRBs 061201 and 080905A. If their redshifts are correct, they may potentially suggest a BNS/NSBH dichotomy in the SGRB population, as this represents the most natural way to explain an apparent contrast in the ejected masses available to power a KN; an NSBH merger can produce as much as 10 times more dynamical ejecta than can a BNS \citep{Metzger17}.

The LIGO/Virgo detection of GW~170817 and electromagnetic follow-up and identification of SGRB 170717A and \KN{} brings about the advent of KN astronomy. Further observations of KNe will reveal whether the magnitude of the emission forms a continuum, or continues to display a gap in brightness between two populations. Unsurprisingly, our best constraints come from the SGRBs at the lowest redshifts, and our work emphasises the need to perform KN searches at low $z$ and in nIR filters.

\acknowledgments
We thank the anonymous referee for useful comments that improved the manuscript.

BG, AJL, NRT \& KW have received funding from the European Research Council (ERC) under the European Union's Horizon 2020 research and innovation programme (grant agreement no 725246, TEDE, PI Levan).

AJL, JDL acknowledge support from STFC via grant ST/P000495/1. 

NRT, KW acknowledge support from STFC via grant ST/N000757/1. 

JH was supported by a VILLUM FONDEN Investigator grant (project number 16599).

SC acknowledges partial funding from Agenzia Spaziale Italiana-Istituto Nazionale di Astrofisica grant I/004/11/3.

SRO gratefully acknowledges the support of the Leverhulme Trust Early Career Fellowship.

This work made use of data supplied by the UK Swift Science Data Centre at the University of Leicester.

\bibliographystyle{aasjournal}
\bibliography{ref}

\begin{thebibliography}{}
\expandafter\ifx\csname natexlab\endcsname\relax\def\natexlab#1{#1}\fi
\providecommand{\url}[1]{\href{#1}{#1}}

\bibitem[{{Andreev} {et~al.}(2010){Andreev}, {Sergeev}, {Parakhin}, {Karpov},
  {Kuznietsova}, {Petkov}, \& {Pozanenko}}]{Andreev10}
{Andreev}, M., {Sergeev}, A., {Parakhin}, N., {et~al.} 2010, GRB Coordinates
  Network, Circular Service, No.~10455, \#1 (2010), 10455

\bibitem[{{Bazin} {et~al.}(2011){Bazin}, {Ruhlmann-Kleider},
  {Palanque-Delabrouille}, {Rich}, {Aubourg}, {Astier}, {Balland}, {Basa},
  {Carlberg}, {Conley}, {Fouchez}, {Guy}, {Hardin}, {Hook}, {Howell}, {Pain},
  {Perrett}, {Pritchet}, {Regnault}, {Sullivan}, {Fourmanoit},
  {Gonz{\'a}lez-Gait{\'a}n}, {Lidman}, {Perlmutter}, {Ripoche}, \&
  {Walker}}]{Bazin11}
{Bazin}, G., {Ruhlmann-Kleider}, V., {Palanque-Delabrouille}, N., {et~al.}
  2011, \aap, 534, A43

\bibitem[{{Belczynski} {et~al.}(2006){Belczynski}, {Perna}, {Bulik},
  {Kalogera}, {Ivanova}, \& {Lamb}}]{Belczynski06}
{Belczynski}, K., {Perna}, R., {Bulik}, T., {et~al.} 2006, \apj, 648, 1110

\bibitem[{{Berger}(2006)}]{Berger06}
{Berger}, E. 2006, GRB Coordinates Network, 5952

\bibitem[{{Berger}(2010)}]{Berger10}
---. 2010, \apj, 722, 1946

\bibitem[{{Berger} \& {Boss}(2005)}]{Berger05c}
{Berger}, E., \& {Boss}, A. 2005, GRB Coordinates Network, 4323

\bibitem[{{Berger} {et~al.}(2009){Berger}, {Cenko}, {Fox}, \&
  {Cucchiara}}]{Berger09}
{Berger}, E., {Cenko}, S.~B., {Fox}, D.~B., \& {Cucchiara}, A. 2009, \apj, 704,
  877

\bibitem[{{Berger} \& {Chornock}(2010)}]{Berger10b}
{Berger}, E., \& {Chornock}, R. 2010, GRB Coordinates Network, Circular
  Service, No.~10410, \#1 (2010), 10410

\bibitem[{{Berger} {et~al.}(2013){Berger}, {Fong}, \& {Chornock}}]{Berger13}
{Berger}, E., {Fong}, W., \& {Chornock}, R. 2013, \apjl, 774, L23

\bibitem[{{Berger} {et~al.}(2007){Berger}, {Morrell}, \& {Roth}}]{Berger07b}
{Berger}, E., {Morrell}, N., \& {Roth}, M. 2007, GRB Coordinates Network, 7151

\bibitem[{{Berger} {et~al.}(2005){Berger}, {Price}, {Cenko}, {Gal-Yam},
  {Soderberg}, {Kasliwal}, {Leonard}, {Cameron}, {Frail}, {Kulkarni}, {Murphy},
  {Krzeminski}, {Piran}, {Lee}, {Roth}, {Moon}, {Fox}, {Harrison}, {Persson},
  {Schmidt}, {Penprase}, {Rich}, {Peterson}, \& {Cowie}}]{Berger05b}
{Berger}, E., {Price}, P.~A., {Cenko}, S.~B., {et~al.} 2005, \nat, 438, 988

\bibitem[{{Blanchard} {et~al.}(2017){Blanchard}, {Berger}, {Fong}, {Nicholl},
  {Leja}, {Conroy}, {Alexander}, {Margutti}, {Williams}, {Doctor}, {Chornock},
  {Villar}, {Cowperthwaite}, {Annis}, {Brout}, {Brown}, {Chen}, {Eftekhari},
  {Frieman}, {Holz}, {Metzger}, {Rest}, {Sako}, \&
  {Soares-Santos}}]{Blanchard17}
{Blanchard}, P.~K., {Berger}, E., {Fong}, W., {et~al.} 2017, \apjl, 848, L22

\bibitem[{{Bloom} {et~al.}(2002){Bloom}, {Kulkarni}, \& {Djorgovski}}]{Bloom02}
{Bloom}, J.~S., {Kulkarni}, S.~R., \& {Djorgovski}, S.~G. 2002, \aj, 123, 1111

\bibitem[{{Bloom} {et~al.}(2006){Bloom}, {Prochaska}, {Pooley}, {Blake},
  {Foley}, {Jha}, {Ramirez-Ruiz}, {Granot}, {Filippenko}, {Sigurdsson},
  {Barth}, {Chen}, {Cooper}, {Falco}, {Gal}, {Gerke}, {Gladders}, {Greene},
  {Hennanwi}, {Ho}, {Hurley}, {Koester}, {Li}, {Lubin}, {Newman}, {Perley},
  {Squires}, \& {Wood-Vasey}}]{Bloom06}
{Bloom}, J.~S., {Prochaska}, J.~X., {Pooley}, D., {et~al.} 2006, \apj, 638, 354

\bibitem[{{Blustin} {et~al.}(2005){Blustin}, {Mangano}, {Voges}, {Marshall}, \&
  {Gehrels}}]{Blustin05}
{Blustin}, A.~J., {Mangano}, V., {Voges}, W., {Marshall}, F., \& {Gehrels}, N.
  2005, GRB Coordinates Network, 4331

\bibitem[{{Bolmer} {et~al.}(2017){Bolmer}, {Steinle}, \& {Schady}}]{Bolmer17}
{Bolmer}, J., {Steinle}, H., \& {Schady}, P. 2017, GRB Coordinates Network,
  Circular Service, No.~21050, \#1 (2017), 21050

\bibitem[{{Butler} {et~al.}(2015){Butler}, {Watson}, {Kutyrev}, {Lee},
  {Richer}, {Klein}, {Fox}, {Prochaska}, {Bloom}, {Cucchiara}, {Troja},
  {Littlejohns}, {Ramirez-Ruiz}, {de Diego}, {Georgiev}, {Gonzalez},
  {Roman-Zuniga}, {Gehrels}, {Moseley}, \& {Beardmore}}]{Butler15}
{Butler}, N., {Watson}, A.~M., {Kutyrev}, A., {et~al.} 2015, GRB Coordinates
  Network, Circular Service, No.~17762, \#1 (2015), 17762

\bibitem[{{Castro-Tirado} {et~al.}(2005){Castro-Tirado}, {de Ugarte Postigo},
  {Gorosabel}, {Fathkullin}, {Sokolov}, {Bremer}, {M{\'a}rquez},
  {Mar{\'{\i}}n}, {Guziy}, {Jel{\'{\i}}nek}, {Kub{\'a}nek}, {Hudec}, {Vitek},
  {Mateo Sanguino}, {Eigenbrod}, {P{\'e}rez-Ram{\'{\i}}rez}, {Sota},
  {Masegosa}, {Prada}, \& {Moles}}]{Castro-Tirado05}
{Castro-Tirado}, A.~J., {de Ugarte Postigo}, A., {Gorosabel}, J., {et~al.}
  2005, \aap, 439, L15

\bibitem[{{Cenko} {et~al.}(2006){Cenko}, {Fox}, \& {Price}}]{Cenko06b}
{Cenko}, S.~B., {Fox}, D.~B., \& {Price}, P.~A. 2006, GRB Coordinates Network,
  5912

\bibitem[{{Cenko} {et~al.}(2007){Cenko}, {Rau}, {Berger}, {Price}, \&
  {Cucchiara}}]{Cenko07}
{Cenko}, S.~B., {Rau}, A., {Berger}, E., {Price}, P.~A., \& {Cucchiara}, A.
  2007, GRB Coordinates Network, 6664

\bibitem[{{Chester} \& {D'Elia}(2015)}]{Chester15}
{Chester}, M.~M., \& {D'Elia}, V. 2015, GRB Coordinates Network, Circular
  Service, No.~17323, \#1 (2015), 17323

\bibitem[{{Chornock} \& {Fong}(2015)}]{Chornock15}
{Chornock}, R., \& {Fong}, W. 2015, GRB Coordinates Network, Circular Service,
  No.~17358, \#1 (2015), 17358

\bibitem[{{Chornock} {et~al.}(2017){Chornock}, {Berger}, {Kasen},
  {Cowperthwaite}, {Nicholl}, {Villar}, {Alexander}, {Blanchard}, {Eftekhari},
  {Fong}, {Margutti}, {Williams}, {Annis}, {Brout}, {Brown}, {Chen}, {Drout},
  {Farr}, {Foley}, {Frieman}, {Fryer}, {Herner}, {Holz}, {Kessler}, {Matheson},
  {Metzger}, {Quataert}, {Rest}, {Sako}, {Scolnic}, {Smith}, \&
  {Soares-Santos}}]{Chornock17}
{Chornock}, R., {Berger}, E., {Kasen}, D., {et~al.} 2017, \apjl, 848, L19

\bibitem[{{Coulter}(2017{\natexlab{a}})}]{Coulter17}
{Coulter}, D.~A. e.~a. 2017{\natexlab{a}}, GRB Coordinates Network, 21529

\bibitem[{{Coulter}(2017{\natexlab{b}})}]{Coulter17a}
---. 2017{\natexlab{b}}, Science, doi:10.1126/science.aap9811

\bibitem[{{Covino} {et~al.}(2006){Covino}, {Malesani}, {Israel}, {D'Avanzo},
  {Antonelli}, {Chincarini}, {Fugazza}, {Conciatore}, {Della Valle}, {Fiore},
  {Guetta}, {Hurley}, {Lazzati}, {Stella}, {Tagliaferri}, {Vietri}, {Campana},
  {Burrows}, {D'Elia}, {Filliatre}, {Gehrels}, {Goldoni}, {Melandri},
  {Mereghetti}, {Mirabel}, {Moretti}, {Nousek}, {O'Brien}, {Pellizza}, {Perna},
  {Piranomonte}, {Romano}, \& {Zerbi}}]{Covino06}
{Covino}, S., {Malesani}, D., {Israel}, G.~L., {et~al.} 2006, \aap, 447, L5

\bibitem[{{Cowperthwaite} {et~al.}(2017){Cowperthwaite}, {Berger}, {Villar},
  {Metzger}, {Nicholl}, {Chornock}, {Blanchard}, {Fong}, {Margutti},
  {Soares-Santos}, {Alexander}, {Allam}, {Annis}, {Brout}, {Brown}, {Butler},
  {Chen}, {Diehl}, {Doctor}, {Drout}, {Eftekhari}, {Farr}, {Finley}, {Foley},
  {Frieman}, {Fryer}, {Garc{\'{\i}}a-Bellido}, {Gill}, {Guillochon}, {Herner},
  {Holz}, {Kasen}, {Kessler}, {Marriner}, {Matheson}, {Neilsen}, {Quataert},
  {Palmese}, {Rest}, {Sako}, {Scolnic}, {Smith}, {Tucker}, {Williams},
  {Balbinot}, {Carlin}, {Cook}, {Durret}, {Li}, {Lopes}, {Louren{\c c}o},
  {Marshall}, {Medina}, {Muir}, {Mu{\~n}oz}, {Sauseda}, {Schlegel}, {Secco},
  {Vivas}, {Wester}, {Zenteno}, {Zhang}, {Abbott}, {Banerji}, {Bechtol},
  {Benoit-L{\'e}vy}, {Bertin}, {Buckley-Geer}, {Burke}, {Capozzi}, {Carnero
  Rosell}, {Carrasco Kind}, {Castander}, {Crocce}, {Cunha}, {D'Andrea}, {da
  Costa}, {Davis}, {DePoy}, {Desai}, {Dietrich}, {Drlica-Wagner}, {Eifler},
  {Evrard}, {Fernandez}, {Flaugher}, {Fosalba}, {Gaztanaga}, {Gerdes},
  {Giannantonio}, {Goldstein}, {Gruen}, {Gruendl}, {Gutierrez}, {Honscheid},
  {Jain}, {James}, {Jeltema}, {Johnson}, {Johnson}, {Kent}, {Krause}, {Kron},
  {Kuehn}, {Nuropatkin}, {Lahav}, {Lima}, {Lin}, {Maia}, {March}, {Martini},
  {McMahon}, {Menanteau}, {Miller}, {Miquel}, {Mohr}, {Neilsen}, {Nichol},
  {Ogando}, {Plazas}, {Roe}, {Romer}, {Roodman}, {Rykoff}, {Sanchez},
  {Scarpine}, {Schindler}, {Schubnell}, {Sevilla-Noarbe}, {Smith}, {Smith},
  {Sobreira}, {Suchyta}, {Swanson}, {Tarle}, {Thomas}, {Thomas}, {Troxel},
  {Vikram}, {Walker}, {Wechsler}, {Weller}, {Yanny}, \&
  {Zuntz}}]{Cowperthwaite17}
{Cowperthwaite}, P.~S., {Berger}, E., {Villar}, V.~A., {et~al.} 2017, \apjl,
  848, L17

\bibitem[{{Cucchiara} {et~al.}(2009){Cucchiara}, {Fox}, {Tanvir}, {Berger},
  {Graham}, \& {Levan}}]{Cucchiara09c}
{Cucchiara}, A., {Fox}, D.~B., {Tanvir}, N., {et~al.} 2009, GRB Coordinates
  Network, 9362

\bibitem[{{Cucchiara} {et~al.}(2013){Cucchiara}, {Prochaska}, {Perley},
  {Cenko}, {Werk}, {Cardwell}, {Turner}, {Cao}, {Bloom}, \&
  {Cobb}}]{Cucchiara13b}
{Cucchiara}, A., {Prochaska}, J.~X., {Perley}, D., {et~al.} 2013, \apj, 777, 94

\bibitem[{{Curran} {et~al.}(2010){Curran}, {Evans}, {de Pasquale}, {Page}, \&
  {van der Horst}}]{Curran10}
{Curran}, P.~A., {Evans}, P.~A., {de Pasquale}, M., {Page}, M.~J., \& {van der
  Horst}, A.~J. 2010, \apjl, 716, L135

\bibitem[{{D'Avanzo} {et~al.}(2015){D'Avanzo}, {D'Elia}, {Lorenzi}, {Mainella},
  {Boschin}, {Garcia de Gurtubai Escudero}, \& {Levan}}]{D'Avanzo15}
{D'Avanzo}, P., {D'Elia}, V., {Lorenzi}, V., {et~al.} 2015, GRB Coordinates
  Network, Circular Service, No.~17326, \#1 (2015), 17326

\bibitem[{{D'Avanzo} {et~al.}(2007){D'Avanzo}, {Piranomonte}, {Antonelli},
  {Covino}, {Fugazza}, {Calzoletti}, {Campana}, {Chincarini}, {Conciatore},
  {Cutini}, {D'Elia}, {Dalessio}, {Fiore}, {Goldoni}, {Guetta}, {Guidorzi},
  {Israel}, {Masetti}, {Melandri}, {Meurs}, {Nicastro}, {Palazzi}, {Pian},
  {Stella}, {Stratta}, {Tagliaferri}, {Tosti}, {Testa}, {Vergani}, \&
  {Vitali}}]{D'Avanzo07b}
{D'Avanzo}, P., {Piranomonte}, S., {Antonelli}, L.~A., {et~al.} 2007, GRB
  Coordinates Network, 7149

\bibitem[{{D'Avanzo} {et~al.}(2009){D'Avanzo}, {Malesani}, {Covino},
  {Piranomonte}, {Grazian}, {Fugazza}, {Margutti}, {D'Elia}, {Antonelli},
  {Campana}, {Chincarini}, {Della Valle}, {Fiore}, {Goldoni}, {Mao}, {Perna},
  {Salvaterra}, {Stella}, {Stratta}, \& {Tagliaferri}}]{D'Avanzo09}
{D'Avanzo}, P., {Malesani}, D., {Covino}, S., {et~al.} 2009, \aap, 498, 711

\bibitem[{{de Pasquale} \& {D'Ai}(2016)}]{dePasquale16}
{de Pasquale}, M., \& {D'Ai}, A. 2016, GRB Coordinates Network, Circular
  Service, No.~19576, \#1 (2016), 19576

\bibitem[{{de Ugarte Postigo} {et~al.}(2014){de Ugarte Postigo}, {Th{\"o}ne},
  {Rowlinson}, {Garc{\'{\i}}a-Benito}, {Levan}, {Gorosabel}, {Goldoni},
  {Schulze}, {Zafar}, {Wiersema}, {S{\'a}nchez-Ram{\'{\i}}rez}, {Melandri},
  {D'Avanzo}, {Oates}, {D'Elia}, {De Pasquale}, {Kr{\"u}hler}, {van der Horst},
  {Xu}, {Watson}, {Piranomonte}, {Vergani}, {Milvang-Jensen}, {Kaper},
  {Malesani}, {Fynbo}, {Cano}, {Covino}, {Flores}, {Greiss}, {Hammer},
  {Hartoog}, {Hellmich}, {Heuser}, {Hjorth}, {Jakobsson}, {Mottola}, {Sparre},
  {Sollerman}, {Tagliaferri}, {Tanvir}, {Vestergaard}, \&
  {Wijers}}]{deUgartePostigo14}
{de Ugarte Postigo}, A., {Th{\"o}ne}, C.~C., {Rowlinson}, A., {et~al.} 2014,
  \aap, 563, A62

\bibitem[{{Eichler} {et~al.}(1989){Eichler}, {Livio}, {Piran}, \&
  {Schramm}}]{Eichler89}
{Eichler}, D., {Livio}, M., {Piran}, T., \& {Schramm}, D.~N. 1989, \nat, 340,
  126

\bibitem[{{Evans} {et~al.}(2017){Evans}, {Cenko}, {Kennea}, {Emery}, {Kuin}, \&
  {Korobkin}}]{Evans17}
{Evans}, P.~A., {Cenko}, S.~B., {Kennea}, J.~A., {et~al.} 2017,
  doi:10.1126/science.aap9580

\bibitem[{{Evans} {et~al.}(2007){Evans}, {Beardmore}, {Page}, {Tyler},
  {Osborne}, {Goad}, {O'Brien}, {Vetere}, {Racusin}, {Morris}, {Burrows},
  {Capalbi}, {Perri}, {Gehrels}, \& {Romano}}]{Evans07}
{Evans}, P.~A., {Beardmore}, A.~P., {Page}, K.~L., {et~al.} 2007, \aap, 469,
  379

\bibitem[{{Evans} {et~al.}(2009){Evans}, {Beardmore}, {Page}, {Osborne},
  {O'Brien}, {Willingale}, {Starling}, {Burrows}, {Godet}, {Vetere}, {Racusin},
  {Goad}, {Wiersema}, {Angelini}, {Capalbi}, {Chincarini}, {Gehrels}, {Kennea},
  {Margutti}, {Morris}, {Mountford}, {Pagani}, {Perri}, {Romano}, \&
  {Tanvir}}]{Evans09}
---. 2009, \mnras, 397, 1177

\bibitem[{{Fermi-GBM}(2017)}]{Fermi17}
{Fermi-GBM}. 2017, GRB Coordinates Network, 524666471

\bibitem[{{Fong} {et~al.}(2015){Fong}, {Berger}, {Margutti}, \&
  {Zauderer}}]{Fong15}
{Fong}, W., {Berger}, E., {Margutti}, R., \& {Zauderer}, B.~A. 2015, \apj, 815,
  102

\bibitem[{{Fong} {et~al.}(2013){Fong}, {Berger}, {Chornock}, {Margutti},
  {Levan}, {Tanvir}, {Tunnicliffe}, {Czekala}, {Fox}, {Perley}, {Cenko},
  {Zauderer}, {Laskar}, {Persson}, {Monson}, {Kelson}, {Birk}, {Murphy},
  {Servillat}, \& {Anglada}}]{Fong13}
{Fong}, W., {Berger}, E., {Chornock}, R., {et~al.} 2013, \apj, 769, 56

\bibitem[{{Fong} {et~al.}(2014){Fong}, {Berger}, {Metzger}, {Margutti},
  {Chornock}, {Migliori}, {Foley}, {Zauderer}, {Lunnan}, {Laskar}, {Desch},
  {Meech}, {Sonnett}, {Dickey}, {Hedlund}, \& {Harding}}]{Fong14}
{Fong}, W., {Berger}, E., {Metzger}, B.~D., {et~al.} 2014, \apj, 780, 118

\bibitem[{{Fong} {et~al.}(2016){Fong}, {Margutti}, {Chornock}, {Berger},
  {Shappee}, {Levan}, {Tanvir}, {Smith}, {Milne}, {Laskar}, {Fox}, {Lunnan},
  {Blanchard}, {Hjorth}, {Wiersema}, {van der Horst}, \& {Zaritsky}}]{Fong16}
{Fong}, W., {Margutti}, R., {Chornock}, R., {et~al.} 2016, \apj, 833, 151

\bibitem[{{Fong} {et~al.}(2017){Fong}, {Berger}, {Blanchard}, {Margutti},
  {Cowperthwaite}, {Chornock}, {Alexander}, {Metzger}, {Villar}, {Nicholl},
  {Eftekhari}, {Williams}, {Annis}, {Brout}, {Brown}, {Chen}, {Doctor},
  {Diehl}, {Holz}, {Rest}, {Sako}, \& {Soares-Santos}}]{Fong17}
{Fong}, W., {Berger}, E., {Blanchard}, P.~K., {et~al.} 2017, \apjl, 848, L23

\bibitem[{{Fox} {et~al.}(2005){Fox}, {Frail}, {Price}, {Kulkarni}, {Berger},
  {Piran}, {Soderberg}, {Cenko}, {Cameron}, {Gal-Yam}, {Kasliwal}, {Moon},
  {Harrison}, {Nakar}, {Schmidt}, {Penprase}, {Chevalier}, {Kumar}, {Roth},
  {Watson}, {Lee}, {Shectman}, {Phillips}, {Roth}, {McCarthy}, {Rauch},
  {Cowie}, {Peterson}, {Rich}, {Kawai}, {Aoki}, {Kosugi}, {Totani}, {Park},
  {MacFadyen}, \& {Hurley}}]{Fox05}
{Fox}, D.~B., {Frail}, D.~A., {Price}, P.~A., {et~al.} 2005, \nat, 437, 845

\bibitem[{{Freiburghaus} {et~al.}(1999){Freiburghaus}, {Rosswog}, \&
  {Thielemann}}]{Freiburghaus99}
{Freiburghaus}, C., {Rosswog}, S., \& {Thielemann}, F.-K. 1999, \apjl, 525,
  L121

\bibitem[{{Gehrels} {et~al.}(2005){Gehrels}, {Sarazin}, {O'Brien}, {Zhang},
  {Barbier}, {Barthelmy}, {Blustin}, {Burrows}, {Cannizzo}, {Cummings}, {Goad},
  {Holland}, {Hurkett}, {Kennea}, {Levan}, {Markwardt}, {Mason}, {Meszaros},
  {Page}, {Palmer}, {Rol}, {Sakamoto}, {Willingale}, {Angelini}, {Beardmore},
  {Boyd}, {Breeveld}, {Campana}, {Chester}, {Chincarini}, {Cominsky},
  {Cusumano}, {de Pasquale}, {Fenimore}, {Giommi}, {Gronwall}, {Grupe}, {Hill},
  {Hinshaw}, {Hjorth}, {Hullinger}, {Hurley}, {Klose}, {Kobayashi},
  {Kouveliotou}, {Krimm}, {Mangano}, {Marshall}, {McGowan}, {Moretti},
  {Mushotzky}, {Nakazawa}, {Norris}, {Nousek}, {Osborne}, {Page}, {Parsons},
  {Patel}, {Perri}, {Poole}, {Romano}, {Roming}, {Rosen}, {Sato}, {Schady},
  {Smale}, {Sollerman}, {Starling}, {Still}, {Suzuki}, {Tagliaferri},
  {Takahashi}, {Tashiro}, {Tueller}, {Wells}, {White}, \& {Wijers}}]{Gehrels05}
{Gehrels}, N., {Sarazin}, C.~L., {O'Brien}, P.~T., {et~al.} 2005, \nat, 437,
  851

\bibitem[{{Goldstein} {et~al.}(2017){Goldstein}, {Veres}, {Burns}, {Briggs},
  {Hamburg}, {Kocevski}, {Wilson-Hodge}, {Preece}, {Poolakkil}, {Roberts},
  {Hui}, {Connaughton}, {Racusin}, {von Kienlin}, {Dal Canton}, {Christensen},
  {Littenberg}, {Siellez}, {Blackburn}, {Broida}, {Bissaldi}, {Cleveland},
  {Gibby}, {Giles}, {Kippen}, {McBreen}, {McEnery}, {Meegan}, {Paciesas}, \&
  {Stanbro}}]{Goldstein17}
{Goldstein}, A., {Veres}, P., {Burns}, E., {et~al.} 2017, \apjl, 848, L14

\bibitem[{{Gompertz} {et~al.}(2015){Gompertz}, {van der Horst}, {O'Brien},
  {Wynn}, \& {Wiersema}}]{Gompertz15}
{Gompertz}, B.~P., {van der Horst}, A.~J., {O'Brien}, P.~T., {Wynn}, G.~A., \&
  {Wiersema}, K. 2015, \mnras, 448, 629

\bibitem[{{Goriely} {et~al.}(2011){Goriely}, {Bauswein}, \&
  {Janka}}]{Goriely11}
{Goriely}, S., {Bauswein}, A., \& {Janka}, H.-T. 2011, \apjl, 738, L32

\bibitem[{{Gottlieb} {et~al.}(2018){Gottlieb}, {Nakar}, \&
  {Piran}}]{Gottlieb18}
{Gottlieb}, O., {Nakar}, E., \& {Piran}, T. 2018, \mnras, 473, 576

\bibitem[{{Grossman} {et~al.}(2014){Grossman}, {Korobkin}, {Rosswog}, \&
  {Piran}}]{Grossman14}
{Grossman}, D., {Korobkin}, O., {Rosswog}, S., \& {Piran}, T. 2014, \mnras,
  439, 757

\bibitem[{{Grupe} {et~al.}(2006){Grupe}, {Burrows}, {Patel}, {Kouveliotou},
  {Zhang}, {M{\'e}sz{\'a}ros}, {Wijers}, \& {Gehrels}}]{Grupe06}
{Grupe}, D., {Burrows}, D.~N., {Patel}, S.~K., {et~al.} 2006, \apj, 653, 462

\bibitem[{{Haggard} {et~al.}(2017){Haggard}, {Nynka}, {Ruan}, {Kalogera},
  {Cenko}, {Evans}, \& {Kennea}}]{Haggard17}
{Haggard}, D., {Nynka}, M., {Ruan}, J.~J., {et~al.} 2017, \apjl, 848, L25

\bibitem[{{Hjorth} {et~al.}(2017){Hjorth}, {Levan}, {Tanvir}, {Lyman},
  {Wojtak}, \& {Schro der}}]{Hjorth17}
{Hjorth}, J., {Levan}, A.~J., {Tanvir}, N.~R., {et~al.} 2017, \apjl,
  doi:10.3847/2041-8213/aa9110

\bibitem[{{Hjorth} {et~al.}(2005{\natexlab{a}}){Hjorth}, {Sollerman},
  {Gorosabel}, {Granot}, {Klose}, {Kouveliotou}, {Melinder}, {Ramirez-Ruiz},
  {Starling}, {Thomsen}, {Andersen}, {Fynbo}, {Jensen}, {Vreeswijk}, {Castro
  Cer{\'o}n}, {Jakobsson}, {Levan}, {Pedersen}, {Rhoads}, {Tanvir}, {Watson},
  \& {Wijers}}]{Hjorth05b}
{Hjorth}, J., {Sollerman}, J., {Gorosabel}, J., {et~al.} 2005{\natexlab{a}},
  \apjl, 630, L117

\bibitem[{{Hjorth} {et~al.}(2005{\natexlab{b}}){Hjorth}, {Watson}, {Fynbo},
  {Price}, {Jensen}, {J{\o}rgensen}, {Kubas}, {Gorosabel}, {Jakobsson},
  {Sollerman}, {Pedersen}, \& {Kouveliotou}}]{Hjorth05}
{Hjorth}, J., {Watson}, D., {Fynbo}, J.~P.~U., {et~al.} 2005{\natexlab{b}},
  \nat, 437, 859

\bibitem[{{Jakobsson} {et~al.}(2004){Jakobsson}, {Hjorth}, {Fynbo}, {Watson},
  {Pedersen}, {Bj{\"o}rnsson}, \& {Gorosabel}}]{Jakobsson04}
{Jakobsson}, P., {Hjorth}, J., {Fynbo}, J.~P.~U., {et~al.} 2004, \apjl, 617,
  L21

\bibitem[{{Jin} {et~al.}(2016){Jin}, {Hotokezaka}, {Li}, {Tanaka}, {D'Avanzo},
  {Fan}, {Covino}, {Wei}, \& {Piran}}]{Jin16}
{Jin}, Z.-P., {Hotokezaka}, K., {Li}, X., {et~al.} 2016, Nature Communications,
  7, 12898

\bibitem[{{Jin} {et~al.}(2017){Jin}, {Li}, {Wang}, {Wang}, {He}, {Yuan},
  {Zhang}, {Zou}, {Fan}, \& {Wei}}]{Jin17}
{Jin}, Z.-P., {Li}, X., {Wang}, H., {et~al.} 2017, ArXiv e-prints,
  arXiv:1708.07008

\bibitem[{{Just} {et~al.}(2015){Just}, {Bauswein}, {Pulpillo}, {Goriely}, \&
  {Janka}}]{Just15}
{Just}, O., {Bauswein}, A., {Pulpillo}, R.~A., {Goriely}, S., \& {Janka}, H.-T.
  2015, \mnras, 448, 541

\bibitem[{{Kann} {et~al.}(2015){Kann}, {Tanga}, \& {Greiner}}]{Kann15b}
{Kann}, D.~A., {Tanga}, M., \& {Greiner}, J. 2015, GRB Coordinates Network,
  Circular Service, No.~17757, \#1 (2015), 17757

\bibitem[{{Kasliwal} {et~al.}(2017{\natexlab{a}}){Kasliwal}, {Korobkin}, {Lau},
  {Wollaeger}, \& {Fryer}}]{Kasliwal17}
{Kasliwal}, M.~M., {Korobkin}, O., {Lau}, R.~M., {Wollaeger}, R., \& {Fryer},
  C.~L. 2017{\natexlab{a}}, \apjl, 843, L34

\bibitem[{{Kasliwal} {et~al.}(2017{\natexlab{b}}){Kasliwal}, {Nakar}, {Singer},
  {Kaplan}, {Cook}, {Van Sistine}, {Lau}, {Fremling}, {Gottlieb}, {Jencson},
  {Adams}, {Feindt}, {Hotokezaka}, {Ghosh}, {Perley}, {Yu}, {Piran}, {Allison},
  {Anupama}, {Balasubramanian}, {Bannister}, {Bally}, {Barnes}, {Barway},
  {Bellm}, {Bhalerao}, {Bhattacharya}, {Blagorodnova}, {Bloom}, {Brady},
  {Cannella}, {Chatterjee}, {Cenko}, {Cobb}, {Copperwheat}, {Corsi}, {De},
  {Dobie}, {Emery}, {Evans}, {Fox}, {Frail}, {Frohmaier}, {Goobar}, {Hallinan},
  {Harrison}, {Helou}, {Hinderer}, {Ho}, {Horesh}, {Ip}, {Itoh}, {Kasen},
  {Kim}, {Kuin}, {Kupfer}, {Lynch}, {Madsen}, {Mazzali}, {Miller}, {Mooley},
  {Murphy}, {Ngeow}, {Nichols}, {Nissanke}, {Nugent}, {Ofek}, {Qi}, {Quimby},
  {Rosswog}, {Rusu}, {Sadler}, {Schmidt}, {Sollerman}, {Steele}, {Williamson},
  {Xu}, {Yan}, {Yatsu}, {Zhang}, \& {Zhao}}]{Kasliwal17b}
{Kasliwal}, M.~M., {Nakar}, E., {Singer}, L.~P., {et~al.} 2017{\natexlab{b}},
  Science, 358, 1559

\bibitem[{{Knust} {et~al.}(2017){Knust}, {Greiner}, {van Eerten}, {Schady},
  {Kann}, {Chen}, {Delvaux}, {Graham}, {Klose}, {Kr{\"u}hler}, {McConnell},
  {Nicuesa Guelbenzu}, {Perley}, {Schmidl}, {Schweyer}, {Tanga}, \&
  {Varela}}]{Knust17}
{Knust}, F., {Greiner}, J., {van Eerten}, H.~J., {et~al.} 2017, \aap, 607, A84

\bibitem[{{Kocevski} {et~al.}(2010){Kocevski}, {Th{\"o}ne}, {Ramirez-Ruiz},
  {Bloom}, {Granot}, {Butler}, {Perley}, {Modjaz}, {Lee}, {Cobb}, {Levan},
  {Tanvir}, \& {Covino}}]{Kocevski10}
{Kocevski}, D., {Th{\"o}ne}, C.~C., {Ramirez-Ruiz}, E., {et~al.} 2010, \mnras,
  404, 963

\bibitem[{{Kong} {et~al.}(2016){Kong}, {Lee}, {Lin}, {Hou}, \& {Liu}}]{Kong16}
{Kong}, A.~K.~H., {Lee}, M.~Y., {Lin}, Y.-M., {Hou}, X., \& {Liu}, C.~Y. 2016,
  GRB Coordinates Network, Circular Service, No.~19575, \#1 (2016), 19575

\bibitem[{{Korobkin} {et~al.}(2012){Korobkin}, {Rosswog}, {Arcones}, \&
  {Winteler}}]{Korobkin12}
{Korobkin}, O., {Rosswog}, S., {Arcones}, A., \& {Winteler}, C. 2012, \mnras,
  426, 1940

\bibitem[{{Kouveliotou} {et~al.}(1993){Kouveliotou}, {Meegan}, {Fishman},
  {Bhat}, {Briggs}, {Koshut}, {Paciesas}, \& {Pendleton}}]{Kouveliotou93}
{Kouveliotou}, C., {Meegan}, C.~A., {Fishman}, G.~J., {et~al.} 1993, \apjl,
  413, L101

\bibitem[{{Kuin} \& {Beardmore}(2017)}]{Kuin17}
{Kuin}, N.~P.~M., \& {Beardmore}, A.~P. 2017, GRB Coordinates Network, Circular
  Service, No.~21049, \#1 (2017), 21049

\bibitem[{{Kuroda} {et~al.}(2010){Kuroda}, {Yanagisawa}, {Shimizu}, {Nagayama},
  {Toda}, {Yoshida}, \& {Kawai}}]{Kuroda10}
{Kuroda}, D., {Yanagisawa}, K., {Shimizu}, Y., {et~al.} 2010, GRB Coordinates
  Network, Circular Service, No.~10388, \#1 (2010), 10388

\bibitem[{{Kuroda} {et~al.}(2016){Kuroda}, {Hanayama}, {Miyaji}, {Watanabe},
  {Yanagisawa}, {Nagayama}, {Yoshida}, {Ohta}, \& {Kawai}}]{Kuroda16}
{Kuroda}, D., {Hanayama}, H., {Miyaji}, T., {et~al.} 2016, GRB Coordinates
  Network, Circular Service, No.~19571, \#1 (2016), 19571

\bibitem[{{Landsman} \& {Holland}(2010)}]{Landsman10}
{Landsman}, W., \& {Holland}, S. 2010, GRB Coordinates Network, Circular
  Service, No.~10892, \#1 (2010), 10892

\bibitem[{{Lattimer}(2011)}]{Lattimer11}
{Lattimer}, J.~M. 2011, \apss, 336, 67

\bibitem[{{Lattimer} \& {Schramm}(1974)}]{Lattimer74}
{Lattimer}, J.~M., \& {Schramm}, D.~N. 1974, \apjl, 192, L145

\bibitem[{{Lazzati} {et~al.}(2017){Lazzati}, {Perna}, {Morsony},
  {L{\'o}pez-C{\'a}mara}, {Cantiello}, {Ciolfi}, {giacomazzo}, \&
  {Workman}}]{Lazzati17}
{Lazzati}, D., {Perna}, R., {Morsony}, B.~J., {et~al.} 2017, ArXiv e-prints,
  arXiv:1712.03237

\bibitem[{{Leloudas} {et~al.}(2010){Leloudas}, {Xu}, {Malesani}, {Levan},
  {Jakobsson}, \& {Djupvik}}]{Leloudas10}
{Leloudas}, G., {Xu}, D., {Malesani}, D., {et~al.} 2010, GRB Coordinates
  Network, Circular Service, No.~10387, \#1 (2010), 10387

\bibitem[{{Levan} {et~al.}(2007){Levan}, {Jakobsson}, {Hurkett}, {Tanvir},
  {Gorosabel}, {Vreeswijk}, {Rol}, {Chapman}, {Gehrels}, {O'Brien}, {Osborne},
  {Priddey}, {Kouveliotou}, {Starling}, {vanden Berk}, \& {Wiersema}}]{Levan07}
{Levan}, A.~J., {Jakobsson}, P., {Hurkett}, C., {et~al.} 2007, \mnras, 378,
  1439

\bibitem[{{Levan} {et~al.}(2017){Levan}, {Lyman}, {Tanvir}, {Hjorth}, {Mandel},
  {Stanway}, {Steeghs}, {Fruchter}, {Troja}, {Schr{\o}der}, {Wiersema},
  {Bruun}, {Cano}, {Cenko}, {de Ugarte Postigo}, {Evans}, {Fairhurst}, {Fox},
  {Fynbo}, {Gompertz}, {Greiner}, {Im}, {Izzo}, {Jakobsson}, {Kangas},
  {Khandrika}, {Lien}, {Malesani}, {O'Brien}, {Osborne}, {Palazzi}, {Pian},
  {Perley}, {Rosswog}, {Ryan}, {Schulze}, {Sutton}, {Th{\"o}ne}, {Watson}, \&
  {Wijers}}]{Levan17}
{Levan}, A.~J., {Lyman}, J.~D., {Tanvir}, N.~R., {et~al.} 2017, \apjl, 848, L28

\bibitem[{{Li} \& {Paczy{\'n}ski}(1998)}]{Li98}
{Li}, L.-X., \& {Paczy{\'n}ski}, B. 1998, \apjl, 507, L59

\bibitem[{{LIGO Scientific} \& {Virgo Collaboration et al.}(2017)}]{Ligo17a}
{LIGO Scientific}, \& {Virgo Collaboration et al.} 2017, GRB Coordinates
  Network, 21505

\bibitem[{{LIGO Scientific Collaboration} {et~al.}(2017){LIGO Scientific
  Collaboration}, {Virgo Collaboration}, \& {Partner Astronomy
  Groups}}]{Ligo17}
{LIGO Scientific Collaboration}, {Virgo Collaboration}, \& {Partner Astronomy
  Groups}. 2017, doi:10.1103/PhysRevLett.119.161101

\bibitem[{{Lyman} {et~al.}(2018){Lyman}, {Lamb}, {Levan}, {Mandel}, {Tanvir},
  {Kobayashi}, {Gompertz}, {Hjorth}, {Fruchter}, {Kangas}, {Steeghs}, {Steele},
  {Cano}, {Copperwheat}, {Evans}, {Fynbo}, {Gall}, {Im}, {Izzo}, {Jakobsson},
  {Milvang-Jensen}, {O'Brien}, {Osborne}, {Palazzi}, {Perley}, {Pian},
  {Rosswog}, {Rowlinson}, {Schulze}, {Stanway}, {Sutton}, {Th{\"o}ne}, {de
  Ugarte Postigo}, {Watson}, {Wiersema}, \& {Wijers}}]{Lyman18}
{Lyman}, J.~D., {Lamb}, G.~P., {Levan}, A.~J., {et~al.} 2018, ArXiv e-prints,
  arXiv:1801.02669

\bibitem[{{Malesani} {et~al.}(2015){Malesani}, {Xu}, {Watson}, \&
  {Blay}}]{Malesani15}
{Malesani}, D., {Xu}, D., {Watson}, D.~J., \& {Blay}, P. 2015, GRB Coordinates
  Network, Circular Service, No.~17756, \#1 (2015), 17756

\bibitem[{{Malesani} {et~al.}(2007){Malesani}, {Covino}, {D'Avanzo}, {D'Elia},
  {Fugazza}, {Piranomonte}, {Ballo}, {Campana}, {Stella}, {Tagliaferri},
  {Antonelli}, {Chincarini}, {Della Valle}, {Goldoni}, {Guidorzi}, {Israel},
  {Lazzati}, {Melandri}, {Pellizza}, {Romano}, {Stratta}, \&
  {Vergani}}]{Malesani07}
{Malesani}, D., {Covino}, S., {D'Avanzo}, P., {et~al.} 2007, \aap, 473, 77

\bibitem[{{Mandel}(2018)}]{Mandel18}
{Mandel}, I. 2018, \apjl, 853, L12

\bibitem[{{Margutti} {et~al.}(2017){Margutti}, {Berger}, {Fong}, {Guidorzi},
  {Alexander}, {Metzger}, {Blanchard}, {Cowperthwaite}, {Chornock},
  {Eftekhari}, {Nicholl}, {Villar}, {Williams}, {Annis}, {Brown}, {Chen},
  {Doctor}, {Frieman}, {Holz}, {Sako}, \& {Soares-Santos}}]{Margutti17}
{Margutti}, R., {Berger}, E., {Fong}, W., {et~al.} 2017, \apjl, 848, L20

\bibitem[{{Margutti} {et~al.}(2018){Margutti}, {Alexander}, {Xie}, {Sironi},
  {Metzger}, {Kathirgamaraju}, {Fong}, {Blanchard}, {Berger}, {MacFadyen},
  {Giannios}, {Guidorzi}, {Hajela}, {Chornock}, {Cowperthwaite}, {Eftekhari},
  {Nicholl}, {Villar}, {Williams}, \& {Zrake}}]{Margutti18}
{Margutti}, R., {Alexander}, K.~D., {Xie}, X., {et~al.} 2018, ArXiv e-prints,
  arXiv:1801.03531

\bibitem[{{Marshall} \& {Beardmore}(2015)}]{Marshall15}
{Marshall}, F.~E., \& {Beardmore}, A.~P. 2015, GRB Coordinates Network,
  Circular Service, No.~17751, \#1 (2015), 17751

\bibitem[{{Marshall} \& {Krimm}(2010)}]{Marshall10}
{Marshall}, F.~E., \& {Krimm}, H.~A. 2010, GRB Coordinates Network, Circular
  Service, No.~10394, \#1 (2010), 10394

\bibitem[{{McLeod} \& {Williams}(2009)}]{McLeod09}
{McLeod}, B., \& {Williams}, G. 2009, GRB Coordinates Network, 9370

\bibitem[{{Melandri} {et~al.}(2006){Melandri}, {Carter}, {Mundell}, {Gomboc},
  {Guidorzi}, {Steele}, {Mottram}, {Bersier}, {Kobayashi}, {Smith}, {Bode},
  {O'Brien}, {Rol}, \& {Bannister}}]{Melandri06}
{Melandri}, A., {Carter}, D., {Mundell}, C., {et~al.} 2006, GRB Coordinates
  Network, 5920

\bibitem[{{M{\'e}sz{\'a}ros} \& {Rees}(1993)}]{Meszaros93b}
{M{\'e}sz{\'a}ros}, P., \& {Rees}, M.~J. 1993, \apj, 405, 278

\bibitem[{{Metzger}(2017)}]{Metzger17}
{Metzger}, B.~D. 2017, Living Reviews in Relativity, 20, 3

\bibitem[{{Metzger} \& {Berger}(2012)}]{Metzger12}
{Metzger}, B.~D., \& {Berger}, E. 2012, \apj, 746, 48

\bibitem[{{Metzger} {et~al.}(2010){Metzger}, {Mart{\'{\i}}nez-Pinedo},
  {Darbha}, {Quataert}, {Arcones}, {Kasen}, {Thomas}, {Nugent}, {Panov}, \&
  {Zinner}}]{Metzger10b}
{Metzger}, B.~D., {Mart{\'{\i}}nez-Pinedo}, G., {Darbha}, S., {et~al.} 2010,
  \mnras, 406, 2650

\bibitem[{{Mirabal} \& {Halpern}(2006)}]{Mirabal06b}
{Mirabal}, N., \& {Halpern}, J.~P. 2006, GRB Coordinates Network, 5906

\bibitem[{{Mooley} {et~al.}(2018){Mooley}, {Nakar}, {Hotokezaka}, {Hallinan},
  {Corsi}, {Frail}, {Horesh}, {Murphy}, {Lenc}, {Kaplan}, {de}, {Dobie},
  {Chandra}, {Deller}, {Gottlieb}, {Kasliwal}, {Kulkarni}, {Myers}, {Nissanke},
  {Piran}, {Lynch}, {Bhalerao}, {Bourke}, {Bannister}, \& {Singer}}]{Mooley18}
{Mooley}, K.~P., {Nakar}, E., {Hotokezaka}, K., {et~al.} 2018, \nat, 554, 207

\bibitem[{{Morgan} {et~al.}(2009){Morgan}, {Cobb}, {Klein}, \&
  {Bloom}}]{Morgan09}
{Morgan}, A.~N., {Cobb}, B.~E., {Klein}, C., \& {Bloom}, J.~S. 2009, GRB
  Coordinates Network, 9359

\bibitem[{{Naito} {et~al.}(2010){Naito}, {Sako}, {Suzuki}, {Kobara}, {Omori},
  {Nagayama}, {Kurita}, \& {Oi}}]{Naito10}
{Naito}, H., {Sako}, T., {Suzuki}, D., {et~al.} 2010, GRB Coordinates Network,
  Circular Service, No.~10889, \#1 (2010), 10889

\bibitem[{{Nakar}(2007)}]{Nakar07}
{Nakar}, E. 2007, \physrep, 442, 166

\bibitem[{{Nicholl} {et~al.}(2017){Nicholl}, {Berger}, {Kasen}, {Metzger},
  {Elias}, {Brice{\~n}o}, {Alexander}, {Blanchard}, {Chornock},
  {Cowperthwaite}, {Eftekhari}, {Fong}, {Margutti}, {Villar}, {Williams},
  {Brown}, {Annis}, {Bahramian}, {Brout}, {Brown}, {Chen}, {Clemens},
  {Dennihy}, {Dunlap}, {Holz}, {Marchesini}, {Massaro}, {Moskowitz},
  {Pelisoli}, {Rest}, {Ricci}, {Sako}, {Soares-Santos}, \&
  {Strader}}]{Nicholl17}
{Nicholl}, M., {Berger}, E., {Kasen}, D., {et~al.} 2017, \apjl, 848, L18

\bibitem[{{Nicuesa Guelbenzu} {et~al.}(2012){Nicuesa Guelbenzu}, {Klose},
  {Greiner}, {Kann}, {Kr{\"u}hler}, {Rossi}, {Schulze}, {Afonso}, {Elliott},
  {Filgas}, {Hartmann}, {K{\"u}pc{\"u} Yolda{\c s}}, {McBreen}, {Nardini},
  {Olivares E.}, {Rau}, {Schmidl}, {Schady}, {Sudilovsky}, {Updike}, \&
  {Yolda{\c s}}}]{NicuesaGuelbenzu12b}
{Nicuesa Guelbenzu}, A., {Klose}, S., {Greiner}, J., {et~al.} 2012, \aap, 548,
  A101

\bibitem[{{Norris} \& {Bonnell}(2006)}]{Norris06}
{Norris}, J.~P., \& {Bonnell}, J.~T. 2006, \apj, 643, 266

\bibitem[{{Norris} {et~al.}(2010){Norris}, {Gehrels}, \& {Scargle}}]{Norris10}
{Norris}, J.~P., {Gehrels}, N., \& {Scargle}, J.~D. 2010, \apj, 717, 411

\bibitem[{{Nysewander} {et~al.}(2009){Nysewander}, {Fruchter}, \&
  {Pe'er}}]{Nysewander09}
{Nysewander}, M., {Fruchter}, A.~S., \& {Pe'er}, A. 2009, \apj, 701, 824

\bibitem[{{Perley} {et~al.}(2009){Perley}, {Kislak}, \&
  {Ganeshalingam}}]{Perley09c}
{Perley}, D.~A., {Kislak}, M., \& {Ganeshalingam}, M. 2009, GRB Coordinates
  Network, 9372

\bibitem[{{Pian} {et~al.}(2017){Pian}, {D'Avanzo}, {Benetti}, {Branchesi},
  {Brocato}, {Campana}, {Cappellaro}, {Covino}, {D'Elia}, {Fynbo}, {Getman},
  {Ghirlanda}, {Ghisellini}, {Grado}, {Greco}, {Hjorth}, {Kouveliotou},
  {Levan}, {Limatola}, {Malesani}, {Mazzali}, {Melandri}, {M{\o}ller},
  {Nicastro}, {Palazzi}, {Piranomonte}, {Rossi}, {Salafia}, {Selsing},
  {Stratta}, {Tanaka}, {Tanvir}, {Tomasella}, {Watson}, {Yang}, {Amati},
  {Antonelli}, {Ascenzi}, {Bernardini}, {Bo{\"e}r}, {Bufano}, {Bulgarelli},
  {Capaccioli}, {Casella}, {Castro-Tirado}, {Chassande-Mottin}, {Ciolfi},
  {Copperwheat}, {Dadina}, {De Cesare}, {di Paola}, {Fan}, {Gendre},
  {Giuffrida}, {Giunta}, {Hunt}, {Israel}, {Jin}, {Kasliwal}, {Klose}, {Lisi},
  {Longo}, {Maiorano}, {Mapelli}, {Masetti}, {Nava}, {Patricelli}, {Perley},
  {Pescalli}, {Piran}, {Possenti}, {Pulone}, {Razzano}, {Salvaterra},
  {Schipani}, {Spera}, {Stamerra}, {Stella}, {Tagliaferri}, {Testa}, {Troja},
  {Turatto}, {Vergani}, \& {Vergani}}]{Pian17}
{Pian}, E., {D'Avanzo}, P., {Benetti}, S., {et~al.} 2017, \nat, 551, 67

\bibitem[{{Planck Collaboration} {et~al.}(2016){Planck Collaboration}, {Ade},
  {Aghanim}, {Arnaud}, {Ashdown}, {Aumont}, {Baccigalupi}, {Banday},
  {Barreiro}, {Bartlett}, \& et~al.}]{PlanckCollaboration16}
{Planck Collaboration}, {Ade}, P.~A.~R., {Aghanim}, N., {et~al.} 2016, \aap,
  594, A13

\bibitem[{{Poole} \& {Troja}(2006)}]{Poole06}
{Poole}, T.~S., \& {Troja}, E. 2006, GRB Coordinates Network, 5069

\bibitem[{{Price} {et~al.}(2006){Price}, {Berger}, {Fox}, {Cenko}, \&
  {Rau}}]{Price06b}
{Price}, P.~A., {Berger}, E., {Fox}, D.~B., {Cenko}, S.~B., \& {Rau}, A. 2006,
  GRB Coordinates Network, 5077

\bibitem[{{Rosswog}(2005)}]{Rosswog05}
{Rosswog}, S. 2005, \apj, 634, 1202

\bibitem[{{Rosswog} {et~al.}(2003){Rosswog}, {Ramirez-Ruiz}, \&
  {Davies}}]{Rosswog03}
{Rosswog}, S., {Ramirez-Ruiz}, E., \& {Davies}, M.~B. 2003, \mnras, 345, 1077

\bibitem[{{Rosswog} {et~al.}(1998){Rosswog}, {Thielemann}, {Davies}, {Benz}, \&
  {Piran}}]{Rosswog98}
{Rosswog}, S., {Thielemann}, F.~K., {Davies}, M.~B., {Benz}, W., \& {Piran}, T.
  1998, in Nuclear Astrophysics, ed. W.~{Hillebrandt} \& E.~{Muller}, 103

\bibitem[{{Rowlinson} {et~al.}(2010){Rowlinson}, {Wiersema}, {Levan}, {Tanvir},
  {O'Brien}, {Rol}, {Hjorth}, {Th{\"o}ne}, {de Ugarte Postigo}, {Fynbo},
  {Jakobsson}, {Pagani}, \& {Stamatikos}}]{Rowlinson10b}
{Rowlinson}, A., {Wiersema}, K., {Levan}, A.~J., {et~al.} 2010, \mnras, 408,
  383

\bibitem[{{Rumyantsev} {et~al.}(2006){Rumyantsev}, {Karimov}, {Salyamov},
  {Pavlenko}, {Efimov}, {Pozanenko}, \& {Ibrahimov}}]{Rumyantsev06}
{Rumyantsev}, V., {Karimov}, R., {Salyamov}, R., {et~al.} 2006, GRB Coordinates
  Network, 5184

\bibitem[{{Rumyantsev} {et~al.}(2010){Rumyantsev}, {Shakhovkoy}, \&
  {Pozanenko}}]{Rumyantsev10}
{Rumyantsev}, V., {Shakhovkoy}, D., \& {Pozanenko}, A. 2010, GRB Coordinates
  Network, Circular Service, No.~10456, \#1 (2010), 10456

\bibitem[{{Ryan} {et~al.}(2015){Ryan}, {van Eerten}, {MacFadyen}, \&
  {Zhang}}]{Ryan15}
{Ryan}, G., {van Eerten}, H., {MacFadyen}, A., \& {Zhang}, B.-B. 2015, \apj,
  799, 3

\bibitem[{{Sari} {et~al.}(1998){Sari}, {Piran}, \& {Narayan}}]{Sari98}
{Sari}, R., {Piran}, T., \& {Narayan}, R. 1998, \apjl, 497, L17

\bibitem[{{Savchenko} {et~al.}(2017){Savchenko}, {Ferrigno}, {Kuulkers},
  {Author}, \& {Author}}]{Savchenko17}
{Savchenko}, V., {Ferrigno}, C., {Kuulkers}, E., {Author}, A.~N., \& {Author},
  A.~N.and~{Author}, A.~N. 2017, \apj

\bibitem[{{Schlafly} \& {Finkbeiner}(2011)}]{Schlafly11}
{Schlafly}, E.~F., \& {Finkbeiner}, D.~P. 2011, \apj, 737, 103

\bibitem[{{Siegel} \& {Beardmore}(2009)}]{Siegel09}
{Siegel}, M.~H., \& {Beardmore}, A.~P. 2009, GRB Coordinates Network, 9369

\bibitem[{{Smartt} {et~al.}(2017){Smartt}, {Chen}, {Jerkstrand}, {Coughlin},
  {Kankare}, {Sim}, {Fraser}, {Inserra}, {Maguire}, {Chambers}, {Huber},
  {Kr{\"u}hler}, {Leloudas}, {Magee}, {Shingles}, {Smith}, {Young}, {Tonry},
  {Kotak}, {Gal-Yam}, {Lyman}, {Homan}, {Agliozzo}, {Anderson}, {Angus},
  {Ashall}, {Barbarino}, {Bauer}, {Berton}, {Botticella}, {Bulla}, {Bulger},
  {Cannizzaro}, {Cano}, {Cartier}, {Cikota}, {Clark}, {De Cia}, {Della Valle},
  {Denneau}, {Dennefeld}, {Dessart}, {Dimitriadis}, {Elias-Rosa}, {Firth},
  {Flewelling}, {Fl{\"o}rs}, {Franckowiak}, {Frohmaier}, {Galbany},
  {Gonz{\'a}lez-Gait{\'a}n}, {Greiner}, {Gromadzki}, {Guelbenzu},
  {Guti{\'e}rrez}, {Hamanowicz}, {Hanlon}, {Harmanen}, {Heintz}, {Heinze},
  {Hernandez}, {Hodgkin}, {Hook}, {Izzo}, {James}, {Jonker}, {Kerzendorf},
  {Klose}, {Kostrzewa-Rutkowska}, {Kowalski}, {Kromer}, {Kuncarayakti},
  {Lawrence}, {Lowe}, {Magnier}, {Manulis}, {Martin-Carrillo}, {Mattila},
  {McBrien}, {M{\"u}ller}, {Nordin}, {O'Neill}, {Onori}, {Palmerio},
  {Pastorello}, {Patat}, {Pignata}, {Podsiadlowski}, {Pumo}, {Prentice}, {Rau},
  {Razza}, {Rest}, {Reynolds}, {Roy}, {Ruiter}, {Rybicki}, {Salmon}, {Schady},
  {Schultz}, {Schweyer}, {Seitenzahl}, {Smith}, {Sollerman}, {Stalder},
  {Stubbs}, {Sullivan}, {Szegedi}, {Taddia}, {Taubenberger}, {Terreran}, {van
  Soelen}, {Vos}, {Wainscoat}, {Walton}, {Waters}, {Weiland}, {Willman},
  {Wiseman}, {Wright}, {Wyrzykowski}, \& {Yaron}}]{Smartt17}
{Smartt}, S.~J., {Chen}, T.-W., {Jerkstrand}, A., {et~al.} 2017, \nat, 551, 75

\bibitem[{{Soares-Santos} {et~al.}(2017){Soares-Santos}, {Holz}, {Annis},
  {Chornock}, {Herner}, {Berger}, {Brout}, {Chen}, {Kessler}, {Sako}, {Allam},
  {Tucker}, {Butler}, {Palmese}, {Doctor}, {Diehl}, {Frieman}, {Yanny}, {Lin},
  {Scolnic}, {Cowperthwaite}, {Neilsen}, {Marriner}, {Kuropatkin}, {Hartley},
  {Paz-Chinch{\'o}n}, {Alexander}, {Balbinot}, {Blanchard}, {Brown}, {Carlin},
  {Conselice}, {Cook}, {Drlica-Wagner}, {Drout}, {Durret}, {Eftekhari}, {Farr},
  {Finley}, {Foley}, {Fong}, {Fryer}, {Garc{\'{\i}}a-Bellido}, {Gill},
  {Gruendl}, {Hanna}, {Kasen}, {Li}, {Lopes}, {Louren{\c c}o}, {Margutti},
  {Marshall}, {Matheson}, {Medina}, {Metzger}, {Mu{\~n}oz}, {Muir}, {Nicholl},
  {Quataert}, {Rest}, {Sauseda}, {Schlegel}, {Secco}, {Sobreira}, {Stebbins},
  {Villar}, {Vivas}, {Walker}, {Wester}, {Williams}, {Zenteno}, {Zhang},
  {Abbott}, {Abdalla}, {Banerji}, {Bechtol}, {Benoit-L{\'e}vy}, {Bertin},
  {Brooks}, {Buckley-Geer}, {Burke}, {Carnero Rosell}, {Carrasco Kind},
  {Carretero}, {Castander}, {Crocce}, {Cunha}, {D'Andrea}, {da Costa}, {Davis},
  {Desai}, {Dietrich}, {Doel}, {Eifler}, {Fernandez}, {Flaugher}, {Fosalba},
  {Gaztanaga}, {Gerdes}, {Giannantonio}, {Goldstein}, {Gruen}, {Gschwend},
  {Gutierrez}, {Honscheid}, {Jain}, {James}, {Jeltema}, {Johnson}, {Johnson},
  {Kent}, {Krause}, {Kron}, {Kuehn}, {Kuhlmann}, {Lahav}, {Lima}, {Maia},
  {March}, {McMahon}, {Menanteau}, {Miquel}, {Mohr}, {Nichol}, {Nord},
  {Ogando}, {Petravick}, {Plazas}, {Romer}, {Roodman}, {Rykoff}, {Sanchez},
  {Scarpine}, {Schubnell}, {Sevilla-Noarbe}, {Smith}, {Smith}, {Suchyta},
  {Swanson}, {Tarle}, {Thomas}, {Thomas}, {Troxel}, {Vikram}, {Wechsler},
  {Weller}, {Dark Energy Survey}, \& {Dark Energy Camera GW-EM
  Collaboration}}]{Soares-Santos17}
{Soares-Santos}, M., {Holz}, D.~E., {Annis}, J., {et~al.} 2017, \apjl, 848, L16

\bibitem[{{Stratta} {et~al.}(2007){Stratta}, {D'Avanzo}, {Piranomonte},
  {Cutini}, {Preger}, {Perri}, {Conciatore}, {Covino}, {Stella}, {Guetta},
  {Marshall}, {Holland}, {Stamatikos}, {Guidorzi}, {Mangano}, {Antonelli},
  {Burrows}, {Campana}, {Capalbi}, {Chincarini}, {Cusumano}, {D'Elia}, {Evans},
  {Fiore}, {Fugazza}, {Giommi}, {Osborne}, {La Parola}, {Mineo}, {Moretti},
  {Page}, {Romano}, \& {Tagliaferri}}]{Stratta07}
{Stratta}, G., {D'Avanzo}, P., {Piranomonte}, S., {et~al.} 2007, \aap, 474, 827

\bibitem[{{Tanvir} {et~al.}(2013){Tanvir}, {Levan}, {Fruchter}, {Hjorth},
  {Hounsell}, {Wiersema}, \& {Tunnicliffe}}]{Tanvir13}
{Tanvir}, N.~R., {Levan}, A.~J., {Fruchter}, A.~S., {et~al.} 2013, \nat, 500,
  547

\bibitem[{{Tanvir} {et~al.}(2015){Tanvir}, {Levan}, {Fruchter}, {Hjorth},
  {Watson}, {Perley}, {Greiner}, {de Ugarte Postigo}, {Thoene}, {Hounsell}, \&
  {Rosswog}}]{Tanvir15}
---. 2015, GRB Coordinates Network, Circular Service, No.~18100, \#1 (2015),
  18100

\bibitem[{{Tanvir} {et~al.}(2017){Tanvir}, {Levan},
  {Gonz{\'a}lez-Fern{\'a}ndez}, {Korobkin}, {Mandel}, {Rosswog}, {Hjorth},
  {D'Avanzo}, {Fruchter}, {Fryer}, {Kangas}, {Milvang-Jensen}, {Rosetti},
  {Steeghs}, {Wollaeger}, {Cano}, {Copperwheat}, {Covino}, {D'Elia}, {de Ugarte
  Postigo}, {Evans}, {Even}, {Fairhurst}, {Figuera Jaimes}, {Fontes}, {Fujii},
  {Fynbo}, {Gompertz}, {Greiner}, {Hodosan}, {Irwin}, {Jakobsson},
  {J{\o}rgensen}, {Kann}, {Lyman}, {Malesani}, {McMahon}, {Melandri},
  {O'Brien}, {Osborne}, {Palazzi}, {Perley}, {Pian}, {Piranomonte}, {Rabus},
  {Rol}, {Rowlinson}, {Schulze}, {Sutton}, {Th{\"o}ne}, {Ulaczyk}, {Watson},
  {Wiersema}, \& {Wijers}}]{Tanvir17}
{Tanvir}, N.~R., {Levan}, A.~J., {Gonz{\'a}lez-Fern{\'a}ndez}, C., {et~al.}
  2017, \apjl, 848, L27

\bibitem[{{Tauris} {et~al.}(2017){Tauris}, {Kramer}, {Freire}, {Wex}, {Janka},
  {Langer}, {Podsiadlowski}, {Bozzo}, {Chaty}, {Kruckow}, {van den Heuvel},
  {Antoniadis}, {Breton}, \& {Champion}}]{Tauris17}
{Tauris}, T.~M., {Kramer}, M., {Freire}, P.~C.~C., {et~al.} 2017, \apj, 846,
  170

\bibitem[{{Thielemann} {et~al.}(2011){Thielemann}, {Arcones}, {K{\"a}ppeli},
  {Liebend{\"o}rfer}, {Rauscher}, {Winteler}, {Fr{\"o}hlich}, {Dillmann},
  {Fischer}, {Martinez-Pinedo}, {Langanke}, {Farouqi}, {Kratz}, {Panov}, \&
  {Korneev}}]{Thielemann11}
{Thielemann}, F.-K., {Arcones}, A., {K{\"a}ppeli}, R., {et~al.} 2011, Progress
  in Particle and Nuclear Physics, 66, 346

\bibitem[{{Troja} {et~al.}(2016{\natexlab{a}}){Troja}, {Sakamoto}, {Cenko},
  {Lien}, {Gehrels}, {Castro-Tirado}, {Ricci}, {Capone}, {Toy}, {Kutyrev},
  {Kawai}, {Cucchiara}, {Fruchter}, {Gorosabel}, {Jeong}, {Levan}, {Perley},
  {Sanchez-Ramirez}, {Tanvir}, \& {Veilleux}}]{Troja16}
{Troja}, E., {Sakamoto}, T., {Cenko}, S.~B., {et~al.} 2016{\natexlab{a}}, \apj,
  827, 102

\bibitem[{{Troja} {et~al.}(2016{\natexlab{b}}){Troja}, {Tanvir}, {Cenko},
  {Levan}, {Barnes}, {Castro-Tirado}, {Fruchter}, {Gehrels}, {Greiner},
  {Kawai}, {Hounsell}, {Hjorth}, {Lien}, {Metzger}, {Perley}, {Rosswog},
  {Sakamoto}, {Thoene}, {de Ugarte Postigo}, \& {Watson}}]{Troja16b}
{Troja}, E., {Tanvir}, N., {Cenko}, S.~B., {et~al.} 2016{\natexlab{b}}, GRB
  Coordinates Network, Circular Service, No.~20222, \#1 (2016), 20222

\bibitem[{{Troja} {et~al.}(2017){Troja}, {Butler}, {Watson}, {Kutyrev}, {Lee},
  {Richer}, {Fox}, {Prochaska}, {Bloom}, {Cucchiara}, {Littlejohns},
  {Ramirez-Ruiz}, {Gonzalez}, {Roman-Zuniga}, {Moseley}, {Capone}, {Golkhou},
  \& {Toy}}]{Troja17}
{Troja}, E., {Butler}, N., {Watson}, A.~M., {et~al.} 2017, GRB Coordinates
  Network, Circular Service, No.~21051, \#1 (2017), 21051

\bibitem[{{Tunnicliffe} {et~al.}(2014){Tunnicliffe}, {Levan}, {Tanvir},
  {Rowlinson}, {Perley}, {Bloom}, {Cenko}, {O'Brien}, {Cobb}, {Wiersema},
  {Malesani}, {de Ugarte Postigo}, {Hjorth}, {Fynbo}, \&
  {Jakobsson}}]{Tunnicliffe14}
{Tunnicliffe}, R.~L., {Levan}, A.~J., {Tanvir}, N.~R., {et~al.} 2014, \mnras,
  437, 1495

\bibitem[{{Updike} {et~al.}(2009){Updike}, {Bryngelson}, \&
  {Milne}}]{Updike09c}
{Updike}, A.~C., {Bryngelson}, G., \& {Milne}, P.~A. 2009, GRB Coordinates
  Network, 9361

\bibitem[{{Villar} {et~al.}(2017){Villar}, {Guillochon}, {Berger}, {Metzger},
  {Cowperthwaite}, {Nicholl}, {Alexander}, {Blanchard}, {Chornock},
  {Eftekhari}, {Fong}, {Margutti}, \& {Williams}}]{Villar17}
{Villar}, V.~A., {Guillochon}, J., {Berger}, E., {et~al.} 2017, \apjl, 851, L21

\bibitem[{{von Kienlin} {et~al.}(2017){von Kienlin}, {Meegan}, {Goldstein},
  {Author}, {Author}, \& {Author}}]{vonKienlin17}
{von Kienlin}, A., {Meegan}, C., {Goldstein}, A., {et~al.} 2017, GRB
  Coordinates Network, 21520

\bibitem[{{Yang} {et~al.}(2015){Yang}, {Jin}, {Li}, {Covino}, {Zheng},
  {Hotokezaka}, {Fan}, {Piran}, \& {Wei}}]{Yang15}
{Yang}, B., {Jin}, Z.-P., {Li}, X., {et~al.} 2015, Nature Communications, 6,
  7323

\end{thebibliography}

\end{document}